\documentclass{elsarticle}
\usepackage{amsmath}
\usepackage{multirow}
\usepackage[table]{xcolor}
\usepackage{hyperref}

\usepackage{tabulary}
\usepackage{amssymb}
\usepackage{bm}
\usepackage{graphicx}
\usepackage{wasysym}
\usepackage{url}
\usepackage{algorithm}
\usepackage{algorithmic}
\usepackage{xcolor}
\usepackage{color}
\usepackage{subfig}
\usepackage{enumitem}
\usepackage{mathtools}
\usepackage{comment}

\newcommand{\anbxj}{\textit{AnBxJ}}

\newcommand{\execnarr}{\textit{TypedExecNarr}}

\usepackage{boxedminipage}

\usepackage{ifpdf}
\usepackage{ifthen}

\ifpdf
  \newcommand{\inputfig}[1]{
  }
\else
  \newcommand{\inputfig}[1]{%
\fi

\newcommand{\tsigma}{\mathcal{T}_{\Sigma}}
\newcommand{\tsigmav}{\tsigma(\Var)}
\newcommand{\cryptn}[2]{\{#2\}_{#1}}
\newcommand{\sign}[2]{\mathsf{sign}_{#1}(#2)}
\newcommand{\scrypt}[2]{\{\!| #2 |\!\}_{#1}}
\newcommand{\inv}[1]{\iffont{inv}(#1)}
\newcommand{\invNA}{\iffont{inv}}
\newcommand{\pair}[2]{\langle #1, #2 \rangle}
\newcommand{\fst}[1]{\pi_1(#1)}
\newcommand{\snd}[1]{\pi_2(#1)}

\newcommand{\idfont}[1]{\mathit{#1}}
\newcommand{\HN}{\idfont{HN}}
\newcommand{\NA}{\idfont{NA}}
\newcommand{\NB}{\idfont{NB}}
\newcommand{\na}{\idfont{na}}
\newcommand{\nb}{\idfont{nb}}
\newcommand{\GX}{\idfont{GX}}
\newcommand{\GY}{\idfont{GY}}

\newcommand{\pkNA}{\iffont{pk}}
\newcommand{\skNA}{\iffont{sk}}
\newcommand{\pk}[1]{\pkNA(#1)}
\newcommand{\sk}[1]{\skNA(#1)}
\newcommand{\pkprime}[1]{\iffont{pk'}(#1)}

\newcommand{\senc}{\iffont{enc}}
\newcommand{\sdec}{\iffont{dec}}

\newcommand{\ckNA}{\mathsf{pk}}
\newcommand{\akNA}{\mathsf{sk}}
\newcommand{\ak}[1]{\akNA(#1)}
\newcommand{\ck}[1]{\ckNA(#1)}

\newcommand{\dyS}{\mathcal{DY}}
\newcommand{\dy}[1]{\dyS(#1)}
\newcommand{\dym}[2]{\dyS_{#1}(#2)}
\newcommand{\dymS}[1]{\dyS_{#1}}
\newcommand{\public}{\Sigma_p}

\newcommand{\state}[2]{\stateNA_{#1}(#2)}
\newcommand{\stateNA}{\iffont{state}}
\newcommand{\iknows}[1]{\iknowsNA(#1)}
\newcommand{\iknowsNA}{\iffont{iknows}}
\newcommand{\ifnot}[1]{\iffont{not}(#1)}
\newcommand{\ifdot}{{}_{\;\bullet\;}}

\newcommand{\ifarrow}[1][]{
  \ifthenelse{\equal{#1}{}}{
    \Rightarrow
  }{
    =\!\!\![{#1}]\!\!\!\hspace{1pt}\Rightarrow
  }}

\newcommand{\iffont}[1]{\mathsf{#1}}

\newcommand{\roleA}{\iffont{roleA}} %
\newcommand{\roleB}{\iffont{roleB}} %
\newcommand{\roleR}{\iffont{roleR}} %
\newcommand{\roleC}{\iffont{roleC}} %

\newcommand{\secCh}{\mbox{$\,\bullet\!\!\rightarrow\!\!\bullet\,$}}
\newcommand{\athCh}{\mbox{$\,\bullet\!\!\rightarrow$\,}}
\newcommand{\cnfCh}{\mbox{${\rightarrow\!\!\bullet\,}$}}
\newcommand{\secRCh}{\,\bullet\!\!\!\twoheadrightarrow\!\!\bullet\,}
\newcommand{\athRCh}{\,\bullet\!\!\!\twoheadrightarrow}
\newcommand{\insecCh}{\rightarrow}

\newcommand{\fresh}[1]{\stackrel{\mbox{\scriptsize @}\,}{#1}}
\newcommand{\forward}{\looparrowright}
\newcommand{\fforward}{\fresh{\forward}}
\newcommand{\farrow}{\fresh{\rightarrow}}

\newcommand{\onionCh}{\,\bullet[\rightarrow]\bullet\,}

\newcommand{\idmxCh}{\onionCh}

\newcommand{\dotChdot}{\,\bullet\!\!\leftrightarrow\!\!\bullet\,}
\newcommand{\dotCh}{\,\bullet\!\!\leftrightarrow}
\newcommand{\Chdot}{\leftrightarrow\!\!\bullet\,}

\newcommand{\PsecChP}{\,\circ\!\!\!\rightarrow\!\!\circ\,}
\newcommand{\PsecCh}{\,\circ\!\!\!\rightarrow\!\!\bullet\,}
\newcommand{\secChP}{\,\bullet\!\!\!\rightarrow\!\!\circ\,}
\newcommand{\PathCh}{\,\circ\!\!\!\rightarrow}
\newcommand{\cnfChP}{\rightarrow\!\!\circ\,}

\newcommand{\PChP}{\,\circ\!\!\leftrightarrow\!\!\circ\,}
\newcommand{\dotChP}{\,\bullet\!\!\leftrightarrow\!\!\dot\,}
\newcommand{\PChdot}{\,\circ\!\!\leftrightarrow\!\!\bullet\,}
\newcommand{\PCh}{\,\circ\!\!\leftrightarrow}
\newcommand{\ChP}{\leftrightarrow\!\!\circ\,}

\newcommand{\FathChNA}{\iffont{athCh}}
\newcommand{\FcnfChNA}{\iffont{cnfCh}}
\newcommand{\FsecChNA}{\iffont{secCh}}
\newcommand{\FfathChNA}{\iffont{fathCh}}
\newcommand{\FfsecChNA}{\iffont{fsecCh}}
\newcommand{\FsecCh}[4]{\FsecChNA(#1;#2;#3;#4)}
\newcommand{\FathCh}[3]{\FathChNA(#1;#2;#3)}
\newcommand{\FcnfCh}[2]{\FcnfChNA(#1;#2)}
\newcommand{\FfathCh}[4]{\FfathChNA^{#1}(#2;#3;#4)}
\newcommand{\FfsecCh}[5]{\FfsecChNA^{#1}(#2;#3;#4;#5)}

\newcommand{\athIssue}[1]{\iffont{athIssue(#1)}}
\newcommand{\cnfIssue}[1]{\iffont{cnfIssue(#1)}}
\newcommand{\secIssue}[1]{\iffont{secIssue(#1)}}
\newcommand{\athRely}[1]{\iffont{athRely(#1)}}
\newcommand{\cnfRely}[1]{\iffont{cnfRely(#1)}}
\newcommand{\secRely}[1]{\iffont{secRely(#1)}}

\newcommand{\pkEnc}[1]{\mathit{pkEnc}(#1)}
\newcommand{\pkSig}[1]{\mathit{pkSig}(#1)}

\newcommand{\DownGrade}{\mathrm{DownGrade}}
\newcommand{\Combine}{\mathrm{Combine}}
\newcommand{\SymOne}{\mathrm{Sym_1}}
\newcommand{\SymTwo}{\mathrm{Sym_2}}
\newcommand{\SymTre}{\mathrm{Sym_3}}
\newcommand{\CreatePseudo}{\mathrm{CreatePseudo}}
\newcommand{\UsePseudo}{\mathrm{UsePseudo}}
\newcommand{\UseRealName}{\mathrm{UseRealName}}
\newcommand{\PseudoDownGrade}{\mathrm{PseudoDownGrade}}
\newcommand{\AuthPseudo}{\mathrm{AuthPseudo}}
\newcommand{\AthTTP}{\mathrm{AthTTP}}
\newcommand{\CnfTTP}{\mathrm{CnfTTP}}
\newcommand{\SecTTP}{\mathrm{SecTTP}}

\newcommand{\honest}[1]{\mathit{honest}(#1)}
\newcommand{\dishonest}[1]{\iffont{dishonest}(#1)}
\newcommand{\vddash}{\vdash\!\!\!\vdash}
\newcommand{\mmodels}{\models\hspace{-0.3cm}\models}

\newcommand{\mode}[3]{({#1} \,|\, {#2} \,|\, {#3})}

\newcommand{\IK}{\mathit{IK}}
\newcommand{\agent}[1]{\iffont{agent}(#1)}

\newcommand{\Fresh}{\mathit{Fresh}}
\newcommand{\Payload}{\mathit{Payload}}
\newcommand{\Public}{\mathit{Public}}
\newcommand{\Tag}{\mathit{Tag}}
\newcommand{\lift}[1]{\lceil #1\rceil}

\newcommand{\pos}[1]{\mathit{pos}(#1)}
\newcommand{\Var}{\mathcal{V}}

\newcommand{\tagfont}[1]{\mathsf{#1}}
\newcommand{\tSone}{\tagfont{S_1}}
\newcommand{\tStwo}{\tagfont{S_2}}

\newcommand{\optfix}[2]{}
\newcommand{\dyM}[2]{\mathcal{DY}_{#1}(#2)}
\newcommand{\nf}[1]{#1_{\downarrow C/F}}
\newcommand{\pattern}[2]{{\lceil\,#1\,\rceil}_{#2}}
\newcommand{\decryptPat}[2]{{\lceil\!\!\lceil\,#1\,\rceil\!\!\rceil}_{#2}}

\newcommand{\pubNA}{\mathit{pub}}
\newcommand{\responseNA}{\mathit{response}}
\newcommand{\checkRNA}{\mathit{check}}

\newcommand{\pub}[1]{\pubNA(#1)}
\newcommand{\response}[1]{\responseNA(#1)}
\newcommand{\checkR}[1]{\checkRNA(#1)}

\newcommand{\Keyword}[1]{\mathsf{#1}}

\newcommand{\Protocol}[5]{
  \Keyword{Protocol:}~\mathit{#1}\\
  \Keyword{Types:}\\
  #2
  \Keyword{Knowledge:}\\
  #3
  \Keyword{Actions:}\\
  \begin{array}{cllllcl}
  #4
  \end{array}\\
  \Keyword{Goals:}\\
  \begin{array}{ccccl}
  #5
  \end{array}
}
\newcommand{\MidProtocol}[5]{
  \begin{array}{ccccl}
  #4
  \end{array}\\
}

\newcommand{\MSC}[5]{
  \Keyword{Protocol:}~\mathit{#1}\\
  \begin{array}{ccccl}
    #4
  \end{array}
}

\newcommand{\ShortProtocol}[5]{
  \Keyword{Protocol:}~\mathit{#1}\\
  \begin{array}{ccccl}
  #4
\end{array}\\
}
\newcommand{\CompactProtocol}[5]{
  \begin{array}{ccccl}
  #4
  \hline
  #5
  \end{array}\\
}

\newcommand{\ifrule}[3]{#1\\ \ifarrow[#2]\\ #3\\[2ex]}

\newcommand{\subterm}{\sqsubseteq}
\newcommand{\propersubterm}{\sqsubset}
\newcommand{\supterm}{\sqsupseteq}
\newcommand{\propersupterm}{\sqsupset}

\newcommand{\decryptions}[2]{\mathit{decryptions}_{#1}(#2)}
\newcommand{\patternR}[3]{\pattern{#2}{#3}^{#1}}

\newcommand{\instanceof}{\succeq}

\newcommand{\secret}[2]{\iffont{secret}(#1,#2)}

\newcommand{\idemix}{\textsf{Identity Mixer}}

\newcommand{\xor}{\oplus}
\newcommand{\algo}[1]{\ensuremath{\mathsf{#1}}}
\newcommand{\const}[1]{\algo{#1}}
\newcommand{\vari}[1]{\ensuremath{\mathit{#1}}}

\newcommand{\ana}[2]{\mathit{ana}_{#1}(#2)}
\newcommand{\derivations}[3]{\mathit{derivations}_{#1}(#2,#3)}
\newcommand{\dereq}[3]{\mathit{dereq}_{#1}(#2,#3)}
\newcommand{\checks}[3]{\mathit{checks}_{#1}(#2,#3)}
\newcommand{\given}[1]{\textsl{given}(#1)}
\newcommand{\ungive}[1]{{#1}_*}

\definecolor{grey}{rgb}{0.5,0.5,0.5}\newcommand{\grey}{\color{grey}}
\definecolor{red}{rgb}{1,0,0}\newcommand{\red}{\color{red}}
\definecolor{green}{rgb}{0,0.45,0}\newcommand{\green}{\color{green}}
\definecolor{blue}{rgb}{0,0,1}\newcommand{\blue}{\color{blue}}

\newcommand{\rpif}{\mathbb{P}}
\newcommand{\rif}{\mathbb{R}}
\newcommand{\rf}{\mathbb{F}}
\newcommand{\re}{\mathbb{E}}
\newcommand{\md}{\mathbb{M}}
\newcommand{\traces}{\mathbb{T}}
\newcommand{\sigS}{\iffont{pk}(S)}
\newcommand{\send}[2]{\iffont{snd}_{#1}(#2)}
\newcommand{\recv}[2]{\iffont{rcv}_{#1}(#2)}
\newcommand{\trace}{t}
\newcommand{\evs}{\mathit{evs}}
\newcommand{\evt}{\mathit{evt}}
\newcommand{\mkset}[1]{[#1]} %
\newcommand{\player}[1]{\mathit{player}(#1)}
\newcommand{\used}[1]{{\mathit{ used\;#1}}}

\newcommand{\sem}[1]{[\![ #1 ]\!]}

\newcommand{\mX}{\mathcal{X}}

\newcommand{\PVar}{\mathcal{P}}
\newcommand{\SigmaP}{\Sigma_\PVar}
\newcommand{\tsigmap}{\mathcal{T}_{\SigmaP}}
\newcommand{\arity}[1]{\mathit{arity}(#1)}

\newcommand{\Nat}{\mathbb{N}}
\newcommand{\dcrypt}[2]{\cryptn{#1}{#2}^{-1}}
\newcommand{\dscrypt}[2]{\scrypt{#1}{#2}^{-1}}
\newcommand{\ccs}[1]{\mathit{ccs}(#1)}

\newcommand{\interpretation}{\mathcal{I}}
\newcommand{\intruder}{\mathsf{i}}

\newcommand{\know}[1]{\mathit{know}(#1)}
\newcommand{\verify}[1]{\mathit{verify}(#1)}
\newcommand{\true}{\mathit{true}}

\newcommand{\anl}[1]{&&&&\%\mathit{#1}\\}
\newcommand{\REM}[1]{\;\;\Keyword{\#}\;\;\text{#1}}

\newcommand{\StoreOA}[2]{\Keyword{Store}_{#1}(\,#2\,)}
\newcommand{\CheckStoreOA}[2]{\Keyword{CheckStore}_{#1}(\,#2\,)}
\newcommand{\LoadOA}[2]{#2\leftarrow\Keyword{Load}_{#1}}

\newcommand{\Store}[2]{\multicolumn{5}{l}{
    \quad\StoreOA{#1}{#2}}\\}
\newcommand{\Load}[2]{\multicolumn{5}{l}{
    \quad\LoadOA{#1}{#2}}\\}
\newcommand{\CheckStore}[2]{\multicolumn{5}{l}{
    \quad\CheckStoreOA{#1}{#2}}\\}

\newcommand{\FormNym}{\mathsf{FormNym}}
\newcommand{\GrantCred}{\mathsf{GrantCred}}
\newcommand{\VerifyCred}{\mathsf{VerifyCred}}
\newcommand{\VerifyCredOnNym}{\mathsf{VerifyCredOnNym}}
\newcommand{\masecNA}{\iffont{masec}}
\newcommand{\masec}[1]{\masecNA(#1)}
\newcommand{\ptagNA}{\iffont{p}}
\newcommand{\ptag}[1]{\ptagNA(#1)}

\newcommand{\commitNA}{\iffont{commit}}
\newcommand{\commitIINA}{\iffont{commit_2}}
\newcommand{\commitIIINA}{\iffont{commit_3}}

\newcommand{\commit}[1]{\commitNA(#1)}
\newcommand{\commitII}[1]{\commitIINA(#1)}
\newcommand{\commitIII}[1]{\commitIIINA(#1)}

\newcommand{\zkpNA}{\iffont{zkp}}
\newcommand{\user}[1]{\iffont{user}(#1)}
\newcommand{\owner}[1]{\iffont{owner}(#1)}

\newcommand{\Always}{\iffont{Always}}
\newcommand{\Previously}{\iffont{Previously}}

\newcommand{\pending}[1]{\iffont{pending}(#1)}

\newcommand{\pack}[1]{\mathit{pack}(#1)}
\newcommand{\hide}[1]{\mathit{hide}(#1)}

\newcommand{\ToTrusted}[1]{\Action{#1}{\idmxCh}{T}}
\newcommand{\TrustedTo}[1]{\Action{T}{\idmxCh}{#1}}

\newcommand{\registered}[1]{\iffont{registered}(#1)}
\newcommand{\granted}[1]{\iffont{granted}(#1)}

\newcommand{\verified}[1]{\iffont{verified}(#1)}

\newcommand{\dhh}{\mathit{dhh}}
\newcommand{\dhf}{\mathit{dhf}}
\newcommand{\DHtag}[1]{\mathit{tag}(#1)}
\newcommand{\hk}{\mathit{hk}}
\newcommand{\fk}{\mathit{fk}}
\newcommand{\exps}{\mathit{exps}}

\newcommand{\from}[2]{\fromnoarg(#1;#2)}
\newcommand{\fromnoarg}{\mathit{from}}

\renewcommand{\vector}[1]{\overline{#1}}

\newcommand{\vars}[1]{\mathit{vars}(#1)}
\newcommand{\dom}[1]{\mathit{dom}(#1)}

\newcommand{\drightarrow}{\rightarrow\!\!\!\!\!\rightarrow}
\newcommand{\bigsem}[1]{\left[\!\!\left[#1\right]\!\!\right]}

\newcommand{\hornarrow}{\rightarrow}
\newcommand{\iknowsa}[1]{\mathsf{iknows}(#1}
\newcommand{\signa}[1]{\mathsf{sign}_{#1}}
\newcommand{\todo}[1]{\fbox{\textbf{TODO: }#1}}
\newcommand{\valid}{\mathit{valid}}
\newcommand{\revoked}{\textit{revoked}}
\newcommand{\db}{\mathit{db}}
\newcommand{\ring}{\mathit{ring}}
\newcommand{\PK}{\mathit{PK}}
\newcommand{\NPK}{\mathit{NPK}}
\newcommand{\occurs}{\mathit{occurs}}
\newcommand{\timplies}{\mathit{implies}}
\newcommand{\LFP}{\mathit{LFP}}

\newcommand{\Habs}{\hat{\mathfrak{A}}}
\newcommand{\Abs}{\mathfrak{A}}
\newcommand{\Vara}{\Var_{\mathfrak{A}}}
\newcommand{\tabs}{\mathcal{T}_\mathfrak{A}}
\newcommand{\thabs}{\hat{\mathcal{T}_\mathfrak{A}}}
\newcommand{\ITIF}{{\it IF}}
\newcommand{\CryptoIF}{{\it CCM}}
\newcommand{\ChanIF}{{\it ICM}}

\newcommand{\ann}{\textit{ann}}
\newcommand{\atag}{\mathsf{atag}}
\newcommand{\fatag}{\mathsf{fatag}}
\newcommand{\ctag}{\mathsf{ctag}}
\newcommand{\stag}{\mathsf{stag}}
\newcommand{\fstag}{\mathsf{fstag}}
\newcommand{\btag}{\mathsf{plain}}
\newcommand{\ftag}{\mathsf{fresh}}
\newcommand{\fotag}{\mathsf{forw}}
\newcommand{\blind}{\mathsf{blind}}
\newcommand{\verif}[1]{#1}
\newcommand{\seenNA}{\mathsf{seen}}
\newcommand{\seen}[1]{\seenNA(#1)}
\newcommand{\notseen}[1]{\mathsf{fresh}(#1)}

\newcommand{\ttbracket}[1]{\texttt{[}#1\texttt{]}}
\newcommand{\anb}{\textit{AnB}}
\newcommand{\anbx}{\textit{AnBx}}
\newcommand{\IM}{\texttt{IM}}
\newcommand{\CM}{\texttt{CM}}
\newcommand{\trans}[3]{\llbracket#1\rrbracket^{#2}_{#3}}
\newcommand{\enc}[1]{\llbracket #1 \rrbracket}
\newcommand{\itrans}[3]{\llparenthesis #1 \rrparenthesis^{#2}_{#3}}
\newcommand{\after}[2]{#1 \texttt{ after } #2}
\newcommand{\arr}[1]{=\!\![#1]\!\!\Rightarrow}
\newcommand{\x}{{\cal X}}
\newcommand{\y}{{\cal Y}}
\newcommand{\X}{{\cal X}}
\newcommand{\T}{{\cal T}}
\newcommand{\V}{{\cal V}}
\newcommand{\chan}{\textsf{chan}}

\newcommand{\cchannel}{\mathsf{Chan_{CCM}}}
\newcommand{\ichannel}{\mathsf{Chan_{ICM}}}
\newcommand{\mgu}{\textit{mgu}}
\newcommand{\bl}{\mathsf{b}}

\renewcommand{\exp}{\iffont{exp}}

\renewcommand{\dscrypt}[2]{\scrypt{#1}{#2}}
\renewcommand{\dcrypt}[2]{\cryptn{#1}{#2}}

\newcommand{\signed}[2]{\mathsf{S}_{#1}(#2)}
\newcommand{\nonce}[1]{n_{#1}}
\newcommand{\newkey}[1]{\mathsf{new}\enskip{#1}}
\newcommand{\hmac}[2]{\mathsf{hmac}_{#1}(#2)}
\newcommand{\hashsymbol}{\mathcal{H}}
\newcommand{\hash}[1]{\mathsf{hash}(#1)}
\newcommand{\verdigest}[2]{[\textit{#1:#2}]}
\newcommand{\digest}[1]{[\textit{#1}]}

\newcommand{\contains}[2]{\iffont{contains}(#1,#2)}
\newcommand{\error}{\iffont{error}}
\newcommand{\wrequest}[1]{\iffont{wrequest}(#1)}
\newcommand{\request}[1]{\iffont{request}(#1)}
\newcommand{\witness}[1]{\iffont{witness}(#1)}

\newcommand{\maxd}[1]{\mathit{maxd}(#1)}
\newcommand{\sccs}[1]{\mathit{sccs}(#1)}
\newcommand{\nts}[1]{\mathit{#1}}

\newcommand{\channel}[2]{\mathsf{channel}_{#1}(#2)}
\newcommand{\GYS}{\mathit{GYS}}

\newcommand{\ccsF}[1]{\mathit{ccs}_F(#1)}

\newcommand{\olds}[1]{\oldstylenums{#1}}
\newcommand{\oldsb}[1]{{\bfseries\olds{#1}}}
\newcommand{\mnote}[1]{\stepcounter{ncomm}%
\vbox to0pt{\vss\llap{\tiny\oldsb{\arabic{ncomm}}}\vskip6pt}%
\marginpar{\tiny\bf\raggedright%
{\oldsb{\arabic{ncomm}}}.\hskip0.5em#1}}\newcounter{ncomm}

\newcommand{\miknote}[1]{{\red \mnote{{\red m: #1}}}}

\usepackage{tikz}
\usetikzlibrary{arrows.meta}
\usetikzlibrary{shapes,arrows,matrix,backgrounds,shapes,positioning,petri,topaths}
\usetikzlibrary{chains,calc}
\setcounter{secnumdepth}{3}
\usepackage{pgfplots}
\pgfplotsset{compat=1.9}
\usepackage{booktabs}
\usepackage{doi}
\usepackage{pgf-umlsd}
\usepackage{longtable}

\newtheorem{definition}{Definition}

\let\oldReturn\Return
\def\hd{0.5}
\def\vd{0.5}
\newcommand{\tm}[1]{\tikz[overlay, anchor=base] \node[red] (#1) {};}
\usepackage{eqparbox}
\renewcommand{\algorithmiccomment}[1]{\hfill\eqparbox{COMMENT}{\# #1}}

\let\proof\relax
\let\endproof\relax

\long\def\ednote#1#2{\begin{quote}\framebox{\begin{minipage}{0.99\linewidth}\footnotesize #1: #2\end{minipage}}\end{quote}}
\newcommand{\eBMargin}[2]{\marginpar{\raggedright\footnotesize #1: #2}}

\usepackage{listings}
\definecolor{kwcolor}{rgb}{0.13,0.29,0.53}    %
\definecolor{cmcolor}{rgb}{0.0,0.5,0.0}       %
\definecolor{stcolor}{rgb}{0.6,0.0,0.0}       %
\definecolor{sectcolor}{rgb}{0.55,0.27,0.07}  %

\lstdefinelanguage{AnBx}{
	keywords = {
		Agent, Number, SeqNumber, SymmetricKey, PublicKey, PrivateKey,
		Certified, Function, Untyped, Definitions,
		share, agree, insecurelyagree,
		weakly, authenticates, on, secret, between, empty
	},
	morekeywords = [2]{
		Protocol, Types, Knowledge, Actions, Goals
	},
	alsoletter = {:},
	sensitive = true,
	breaklines = true,
	basicstyle = \footnotesize\ttfamily,
	keywordstyle = \bfseries\color{kwcolor},
	keywordstyle = [2]\bfseries\color{sectcolor},
	commentstyle = \color{cmcolor}\itshape,
	morecomment = [l]{\#},
	morestring = [b]",
	stringstyle = \color{stcolor},
	tabsize = 4
}

\lstdefinelanguage{AnB}{
	keywords = {
		Agent, Number, SeqNumber, SymmetricKey, PublicKey, PrivateKey,
		Function, Certified,
		share, agree,
		weakly, authenticates, on, secret, between
	},
	morekeywords = [2]{
		Protocol, Types, Knowledge, Actions, Goals
	},
	sensitive = true,
	breaklines = true,
	basicstyle = \footnotesize\ttfamily,
	keywordstyle = \bfseries\color{kwcolor},
	keywordstyle = [2]\bfseries\color{sectcolor},
	commentstyle = \color{cmcolor}\itshape,
	morecomment = [l]{\#},
	morestring = [b]",
	stringstyle = \color{stcolor},
	tabsize = 2
}

\lstdefinelanguage{OFMCAttackTrace}{
    keywords = {Reached, State:,request,state_rB,state_rA,state_rs,state_rIntr,ATTACK, TRACE:},
    alsoletter=:,
    breaklines=true,
    basicstyle=\footnotesize\ttfamily,%
    keywordstyle=\bfseries,
    commentstyle=\color{green}\itshape,
    morecomment=[l]{\#},
    tabsize=2,     
}

\lstset{
	language=Java,
	basicstyle=\footnotesize\ttfamily,
	commentstyle=\color{pgreen},
	keywordstyle=\color{kwcolor}\bfseries,
	stringstyle=\color{stcolor},
	tabsize=2,           %
	showtabs=false,
	breaklines=true,
	breakindent=0pt,
	framesep=3pt,
	framerule=0.4pt,
	xleftmargin=2pt,
	xrightmargin=2pt,
	showspaces=false,
}

\newif\ifshowfixes \showfixestrue
\newcommand{\fix}[2]{\ifshowfixes {\bf FIX}\footnote{{\bf FIX #1}: #2} \else \relax \fi}
\newcommand{\rulesep}{\unskip\ \vrule\ }

\definecolor{pgreen}{rgb}{0,0.5,0}
\lstset{language=Java,
  basicstyle=\footnotesize\ttfamily,
  commentstyle=\color{pgreen},
}

\usepackage{nomencl}
\makenomenclature
\renewcommand{\nomname}{}
\setlength{\nomitemsep}{0.25mm}

\extrafloats{100}
\usepackage{versions}
\excludeversion{abstractionslearningextra}
\includeversion{SHORT}
\excludeversion{LONG}
\makeatletter
\def\ps@pprintTitle{%
  \let\@oddhead\@empty
  \let\@evenhead\@empty  \let\@oddfoot\@empty
  \let\@evenfoot\@oddfoot
}
\makeatother
\begin{document}
    \title{Global Surveillance Under the Lens:\\Reconstructing Passive Dolev--Yao Intruders}
	\author{Paolo Modesti}
	\ead{p.modesti@tees.ac.uk}
	\address{Teesside University, Middlesbrough, United Kingdom}

\begin{abstract}
Global surveillance systems record large volumes of network traffic for
retrospective analysis, but packet captures alone reveal little about
protocol-level semantics. This paper presents an automated,
specification-guided methodology for reconstructing passive Dolev--Yao
intruder executions from captured network traffic and validating the
reconstructed executions against a formal protocol specification. The
framework takes a PCAP capture, an \anbx{} specification, and the
associated deployment metadata; it extracts Java-serialised objects from
TCP streams, reconstructs protocol sessions and channel-to-step
assignments, and generates a passive intruder that evaluates the
Dolev--Yao-derived checks produced by the \anbx{} compiler. Three
reconstruction challenges are addressed: multiple serialised objects and
object references within a TCP stream, ambiguity in mapping interleaved
connections to concurrent protocol sessions, and semantic validation of
candidate reconstructions. We evaluate the methodology on 121 protocols
from the \anbx{} test suite. For single-session captures, the standard
JDK cryptographic provider recovers 97.8\% of expected objects and
achieves 94.2\% successful reconstruction, where success requires
complete object recovery and a clean passive-intruder execution. For the
112 parallel-capable protocols, the toolchain achieves 89.3\% successful
reconstruction for both two and three concurrent sessions, while
recovering approximately 97.2\% of expected objects. Experiments with
mismatched protocol-role mappings and deliberately mutated traces show
that the generated Dolev--Yao checks respond to semantic
inconsistencies. By making the reconstruction step explicit, the
solution produces protocol-level observations on which assessments of
protocol resilience and the design of defensive countermeasures can be
built.
\end{abstract}

 \maketitle

\section{Introduction}
\label{sec:introduction}

Global surveillance programmes, as disclosed by Edward Snowden in
2013~\cite{snowden2019permanent}, have fundamentally altered our
understanding of privacy and communication security. Snowden's revelations
exposed a vast ecosystem of state surveillance, largely operated by the US
National Security Agency (NSA) and its partners in the Five Eyes
intelligence alliance. These programmes included \textsc{Prism} (collection
from major US technology companies), \textsc{Upstream} (interception on the
internet backbone), and \textsc{XKeyscore} (real-time interception of user
activity). Others, such as \textsc{Bullrun} (undermining encryption
standards), \textsc{Muscular} (infiltrating the private data links between
Google and Yahoo data centres), and \textsc{Quantuminsert} (targeted
exploitation via network packet injection), further demonstrate the scale
and sophistication of these operations. State actors systematically
intercept, store, and analyse private communications, often exploiting
protocol-level vulnerabilities. The threat is not merely theoretical:
surveillance capabilities are used to bypass encryption, harvest metadata,
and retroactively decrypt captured traffic when cryptographic keys become
compromised~\cite{green2015puncturable}. These practices challenge the
confidentiality and integrity of digital communication, with profound
implications for privacy, economic security, and democratic governance. The
economic cost of this threat has been quantified by
Blanco-Romero et al.~\cite{blancoromero2026}, who demonstrate that storing
intercepted ciphertext is economically trivial for state-level actors over
multi-decade timeframes.

Our work contributes a concrete,
automated mechanism for analysing the technical underpinnings of such
surveillance: specifically, enabling systematic assessment of the information that captured protocol executions expose to a passive observer.

\subsection*{The Passive Adversary Problem}

The Dolev--Yao model~\cite{dolev83ieee} is the standard formalism for
analysing cryptographic protocols in the presence of an adversarial
network. In this symbolic model, the intruder controls the communication
medium: it can intercept, modify, replay, and inject messages. This
abstraction has proven effective for discovering logical flaws in protocol
designs, leading to verification tools such as OFMC~\cite{basin2005ofmc},
ProVerif~\cite{blanchet2016modeling}, and
Tamarin~\cite{tamarin2012automated}. However, the traditional Dolev--Yao
intruder is \emph{active} and, therefore, it manipulates traffic.

In contrast, many real-world surveillance scenarios are \emph{passive}. An
adversary silently records network traffic, storing encrypted
communications for later decryption when keys become available, or exploits
weaknesses in key exchange protocols to compromise forward secrecy in
practice~\cite{weakdh15}. This harvest now, decrypt later
strategy~\cite{blancoromero2026} is particularly concerning for protocols
lacking forward secrecy. Despite its practical relevance, the passive
intruder has received comparatively little attention in the formal methods
literature, with notable
exceptions~\cite{DBLP:journals/jar/DelauneJ06,DBLP:journals/iandc/BaudetCK09},
and remains underexplored in the context of network trace analysis.
Existing verification tools do not directly support the analysis of
captured traffic against protocol specifications. Verification tools can
synthesise \emph{symbolic} attacks, but cannot validate whether a concrete
network trace corresponds to a protocol execution or an attack. Network
forensics tools (e.g., Wireshark~\cite{tshark}) display packet contents
but lack the semantic understanding required to determine whether a
captured trace corresponds to a valid protocol run or contains evidence of
a security violation.

\subsection*{Approach and Contributions}

This paper presents an automated, specification-guided methodology for
reconstructing passive Dolev--Yao intruder executions from captured network
traffic. The framework takes as input a PCAP file (captured under a
man-in-the-middle configuration where the intruder passively relays
messages), a protocol specification in the \anbx{}
language~\cite{jisa2016}, and the associated deployment metadata (in our implementation, a configuration file generated by the compiler),
and produces executable Java code that reconstructs the captured trace. The
Dolev--Yao knowledge analysis is performed by the \anbx{} compiler~\cite{anbx2015} at
compilation time and materialised as receive-side checks in the generated
intruder role; the reconstruction stage supplies concrete observations to
these checks, rather than performing symbolic deduction at run time. The
generated checks are consistency conditions on the captured trace, not a
proof of security (Section~\ref{subsec:compiler-pipeline}).

Our methodology extends the \anbx{} compiler~\cite{GARCIA2024100038} with a
reconstruction pipeline (Figure~\ref{fig:pipeline-overview}) that (1)
extracts serialised Java objects from each TCP stream, preserving object
graphs via \texttt{TC\_REFERENCE} markers by reading complete streams; (2)
pairs bidirectional streams into connections, groups connections into
sessions, and enumerates candidate channel assignments when a connected
component contains multiple interleaved sessions; (3) maps extracted
objects to protocol steps via the compiler-generated channel-to-step map,
resolving streams that carry more than one object; (4) generates a passive
intruder role that reads the reconstructed step files and performs the
compiler-derived defensive checks; and (5) scores each candidate
arrangement by the number of clean sessions and passed checks, selecting
the best-scoring one. The intruder does not modify messages; it observes,
validates, and reconstructs them, faithfully modelling a passive adversary.

We evaluate the framework on a test suite of 121 protocols drawn from the
OFMC distribution~\cite{basin2005ofmc} that ships with the \anbx{}
toolchain and from the SET~\cite{DBLP:journals/jar/BellaMP06} and
iKP~\cite{DBLP:journals/jsac/BellareGHHKSTHW00} families. The pipeline
recovers 97.8\% of expected objects at \(N=1\) and achieves 94.2\%
successful reconstruction; on the 112 parallel-capable protocols it
achieves 89.3\% at both \(N=2\) and \(N=3\). Detailed metrics, including a comparison between the JDK and Bouncy Castle providers, are reported in Section~\ref{sec:results}.

The contributions of this paper are fourfold. First, a concrete operational
instantiation of a passive Dolev--Yao intruder in which the compiler's
symbolic knowledge analysis is materialised as executable receive-side
checks over reconstructed observations. Second, a PCAP-to-protocol
reconstruction pipeline that extracts Java-serialised objects, preserves
object-stream semantics, maps objects to protocol steps, and resolves
interleaved sessions through candidate enumeration. Third,
specification-guided candidate selection, in which Dolev--Yao-derived
consistency checks distinguish competing reconstructions of the same
captured traffic. Fourth, an evaluation over 121 protocols assessing behaviour under multiple sessions, alternative cryptographic providers, third-party Java traffic, mismatched protocol-role
mappings, and deliberately mutated traces. The purpose of the surveillance framing is defensive: the framework makes
explicit what a passive adversary can recover from captured traffic, so
that the resulting knowledge can inform the design of protocols that are
more resilient to retrospective analysis. The prototype does not model a
complete intelligence-collection system or infer unknown protocols from
arbitrary traffic; it provides a controlled laboratory model of one stage
of retrospective surveillance analysis, and a basis for assessing which
protocol design choices reduce the information recoverable after
interception.

\subsection*{Outline of the Paper}

The paper is organised as follows. Section~\ref{sec:background} reviews the
Dolev--Yao model, the \anbx{} language, the compiler pipeline, and relevant
background on Java serialisation and PCAP processing.
Section~\ref{sec:methodology} presents the reconstruction workflow from PCAP
to executable Java code. Section~\ref{sec:results} reports the experimental results,
Section~\ref{sec:discussion} discusses them and their limitations, and
Section~\ref{sec:related-work} discusses related work.
Section~\ref{sec:conclusion} concludes and outlines future work.

\section{Background}
\label{sec:background}

This section introduces the essential background for understanding our
passive intruder reconstruction framework. We begin by defining the
Dolev--Yao model with a specific focus on the passive adversary. We then
present the \anbx{} specification language and the compiler pipeline that
transforms abstract protocol descriptions into executable Java code.
Finally, we discuss Java serialisation, the technical foundations that
enable extraction of objects from captured network traffic.

\subsection{The Passive Adversary and Threat Model}
\label{subsec:threat-model}

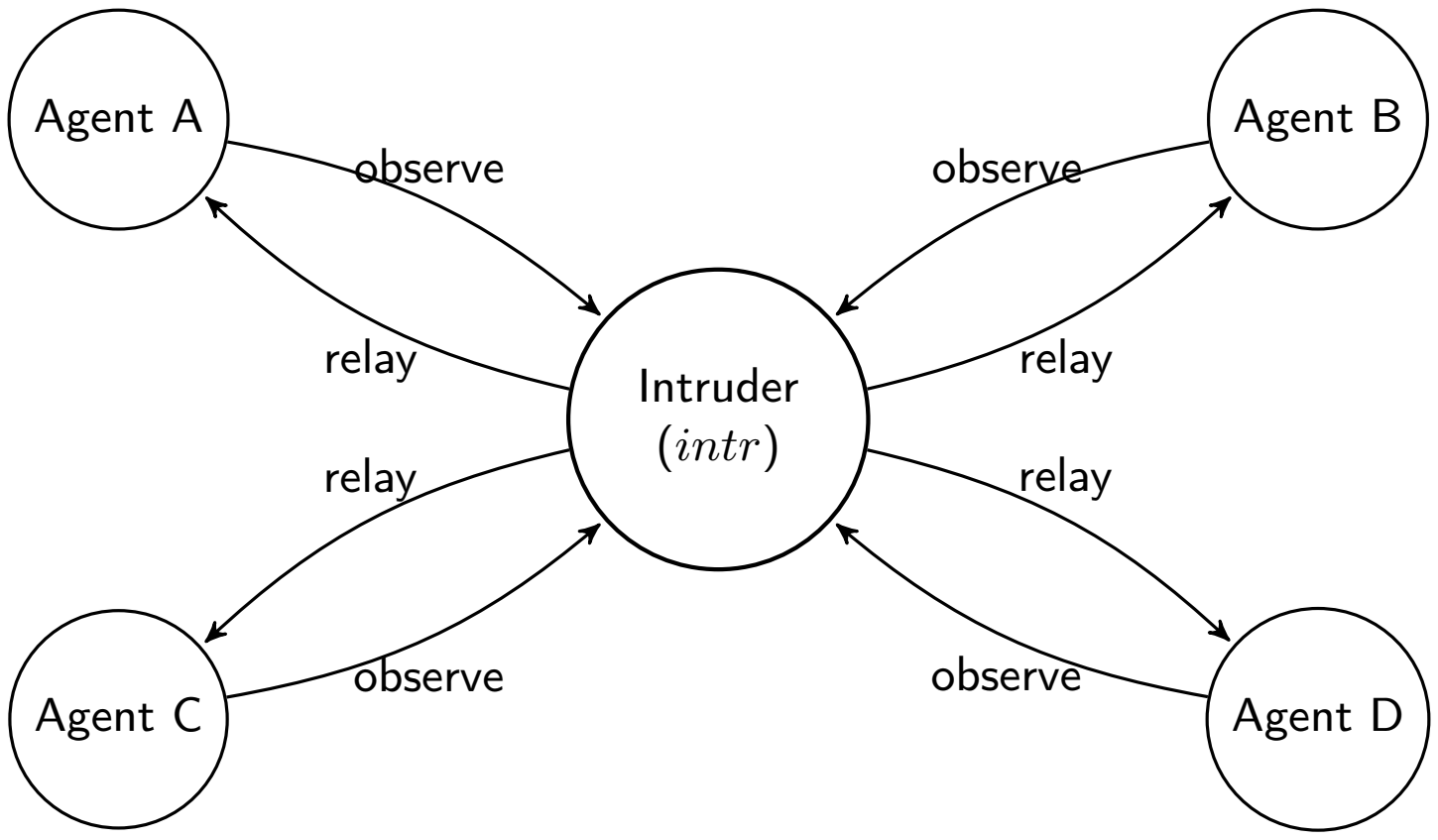
\begin{figure}[t]
	\centering
	\begin{tikzpicture}[
		->,
		>=stealth',
		shorten >=1pt,
		auto,
		node distance=4cm,
		semithick,
		font=\small\sffamily
		]
		\tikzset{
			agent/.style={circle, draw, minimum size=1cm},
			intruder/.style={circle, draw, thick, minimum size=2cm, align=center}
		}
		\node[agent] (A) at (-4,2) {Agent A};
		\node[agent] (B) at (4,2) {Agent B};
		\node[agent] (C) at (-4,-2) {Agent C};
		\node[agent] (D) at (4,-2) {Agent D};
		\node[intruder] (I) at (0,0) {Intruder\\($intr$)};
		\path (A) edge[bend left=15] node[midway,above] {observe} (I);
		\path (I) edge[bend left=15] node[midway,below] {relay} (A);
		\path (B) edge[bend right=15] node[midway,above] {observe} (I);
		\path (I) edge[bend right=15] node[midway,below] {relay} (B);
		\path (C) edge[bend right=15] node[midway,below] {observe} (I);
		\path (I) edge[bend right=15] node[midway,above] {relay} (C);
		\path (D) edge[bend left=15] node[midway,below] {observe} (I);
		\path (I) edge[bend left=15] node[midway,above] {relay} (D);
	\end{tikzpicture}
	\caption{Passive Dolev--Yao adversary model: all communication is
		mediated by the intruder, who observes and transparently forwards
		messages.}  %
	\label{fig:dolev-yao}
\end{figure}

Our work assumes a passive Dolev--Yao adversary~\cite{dolev83ieee} that
observes all network communication between honest agents. In the
man-in-the-middle configuration we consider, messages sent by one honest
agent are received by the intruder and forwarded, unmodified, to their
intended recipient.
Figure~\ref{fig:dolev-yao} illustrates this communication pattern. In the
active model, the intruder may also modify, drop or inject messages; in
the passive model, the intruder's capabilities are restricted to
observation and transparent forwarding.

The passive adversary model captures the threat posed by bulk surveillance
operations that record network traffic for later analysis: the adversary
can observe all messages in transit, store them, and later analyse them
when cryptographic keys become available.

\begin{figure}[ht]
	\centering
	\begin{tabular}{rll>{\raggedright}m{3.3cm}}
		$\ensuremath{\iknows{M}.\iknows{K}}$ & $\ifarrow$ & $\iknows{\cryptn{K}{M}}$ & \multirow{3}{3.3cm}{\emph{Asymmetric}\\
			\emph{Encryption}}\tabularnewline
		\addlinespace
		$\iknows{\cryptn{K}{M}}.\iknows{\inv{K}}$ & $\ifarrow$ & $\iknows{M}$ & \tabularnewline
		\addlinespace
		$\iknows{\cryptn{\inv{K}}{M}}$ & $\ifarrow$ & $\iknows{M}$ & \tabularnewline
		\addlinespace
		$\ensuremath{\iknows{M}.\iknows{K}}$ & $\ifarrow$ & $\iknows{\scrypt{K}{M}}$ & \multirow{2}{3.3cm}{\emph{Symmetric}\\
			\emph{Encryption}}\tabularnewline
		\addlinespace
		$\iknows{\scrypt{K}{M}}.\iknows{K}$ & $\ifarrow$ & $\iknows{M}$ & \tabularnewline
		\addlinespace
		$\iknows{M}.\iknows{N}$ & $\ifarrow$ & $\iknows{M,N}$ & \emph{Tupling}\tabularnewline
		\addlinespace
		$\iknows{M,N}$ & $\ifarrow$ & $\iknows{M}.\iknows{N}$ & \emph{Projection}\tabularnewline
		\addlinespace
		$\iknows{M_{1}}.\cdots.\iknows{M_{n}}$ & $\ifarrow$ & $\iknows{f(M_{1},\ldots,M_{n})}$ & \emph{Function\\Application}\\ %
	\end{tabular}
	\caption{Dolev--Yao intruder deduction rules. The last rule applies
		only to public function symbols; private functions (declared with
		\texttt{->*}) cannot be constructed by the intruder from known
		arguments.}
	\label{fig:Dolev--Yao-intruder-rules}
\end{figure}

Formally, the passive intruder has the same deductive capability as the
active intruder but is restricted in its actions: it observes messages
and may derive new terms from its knowledge, but it cannot construct and
transmit arbitrary messages of its own choosing. The intruder's knowledge
evolves as follows:

\begin{itemize}
	\item Initially, \(\mathcal{K}_0\) contains all public information:
	agent identities, public keys, and protocol-specific public constants.
	\item Upon observing a message \(m\) on the network, the intruder adds
	\(m\) to its knowledge: \(\mathcal{K} := \mathcal{K} \cup \{m\}\).
	\item The intruder can apply the standard Dolev--Yao deduction rules
	(Figure~\ref{fig:Dolev--Yao-intruder-rules}) to derive new terms from
	its knowledge.
\end{itemize}

This deductive capability is crucial for our reconstruction: the intruder
must parse observed messages, verify their well-formedness, and validate
that they are consistent with the protocol specification.

\paragraph{Compiled checks and Dolev--Yao deduction}
The reconstruction stage evaluates the consequences of the compiler's
Dolev--Yao analysis rather than performing symbolic deduction at run
time. The \anbx{} compiler analyses the protocol under the Dolev--Yao
deduction rules at compile time and materialises the consequences of
that analysis as receive-side checks in the generated intruder role.
The checks are derived from the compiler's
analysis~\cite{jisa2016, anbx2015} of the specification, not from the
implementation, and express the constraints the analysis places on
received messages; the analysis has been validated empirically on a
broad test suite of security protocols~\cite{jisa2016}. 
The compiled intruder therefore mirrors a Dolev--Yao intruder whose
deduction has been performed in advance: the checks are the same ones the
compiler would have generated for a network execution, and their outcomes
depend only on the recovered messages and the specification, not on
whether those messages arrived over a socket or were read from a file.

The adversary is assumed to have knowledge of the protocol specification
(in the \anbx{} language), the public keys of all honest agents, the
network configuration (addresses and ports), the mapping from observed IP
addresses and ports to protocol roles, the channel structure, and the
agent-role mapping. It is also assumed to know the wire format used by the
protocol implementation (here, Java serialisation over TCP). Decryption is
only possible if cryptographic keys are public or become available through
compromise or cryptanalysis. From a surveillance perspective, this
information corresponds to what may be obtained from protocol
documentation, network metadata, prior observation, or other intelligence
sources. The contribution is not that a passive adversary can infer all of
this from an arbitrary PCAP; it is that, once this knowledge is available,
the semantic reconstruction itself can be automated. The extraction stage
does not depend on the compiler having produced the implementation, and
the defensive checks are derived from the specification, not from the
implementation; both claims are substantiated in
Sections~\ref{subsubsec:handwritten} and~\ref{subsubsec:discrimination}.

We assume all communication follows the standard Dolev--Yao pattern
\(A \rightarrow intr \rightarrow B\), where the intruder \(intr\) simply
forwards messages without modification. Complete observation of all
messages is assumed; if messages are lost or reordered, the reconstruction
may not correspond to a valid protocol execution.

This passive adversary model defines the threat landscape addressed by our
reconstruction framework: instantiating the passive Dolev--Yao intruder
operationally means reconstructing its observed protocol execution from
the PCAP and evaluating the specification-derived checks against the
recovered messages. The following subsection describes the \anbx{}
specification language, which provides the protocol models used throughout
the paper.

\subsection{Protocol Specification with \anbx{}}
\label{subsec:anbx-language}

The \anbx{} language~\cite{jisa2016} extends the Alice \& Bob notation
(\anb{}~\cite{DBLP:conf/IEEEares/Modersheim09}) with channel abstractions
that allow protocol designers to specify security properties
independently of their concrete cryptographic realisation. While these
abstractions are useful for high-level protocol design, our
reconstruction pipeline operates on the concrete compiled form of the
specification, after channel abstractions have been resolved. This is
because the captured PCAP traffic contains real-world messages, not
abstract channels. We therefore present the language in its concrete
\anb{} form, which is the input to the compiler stages that generate
executable code.

A protocol specification consists of four mandatory sections:

\begin{description}
	\item \textit{Types} Declares agent roles, function symbols, and type
	signatures. Functions may be declared \emph{private} (denoted by
	\(\to^*\)), meaning the intruder cannot construct the function's
	output even if it knows the arguments.
	\item \textit{Knowledge} Defines the initial knowledge of each agent.
	Agents may also \emph{share} terms (e.g., pre-shared symmetric keys) or
	\emph{agree} on values prior to protocol execution.
	\item \textit{Actions} Describes the protocol's message exchanges in
	an ideal execution.
	\item \textit{Goals} Declares the security properties to be verified
	(secrecy, authentication, weak authentication).
\end{description}

The \textit{Actions} section defines the expected sequence of messages,
their ordering, and the channels over which they are transmitted. This
information is used by the compiler to generate the channel-to-step
mapping that guides the reconstruction process described in
Section~\ref{sec:methodology}.

Figure~\ref{fig:anbx-example} illustrates a complete specification of the
Carlsen protocol~\cite{carlsen}, a classic authenticated key exchange
protocol that relies on a trusted server \(s\).

\begin{figure}[ht]
	\begin{lstlisting}[language=AnB]
		Protocol: Carlsen AnB
		
		Types:
		Agent A,B,s;
		Number NxNA,NxNB,NxNB2;
		SymmetricKey KxKAB;
		Function [Agent,Agent ->* SymmetricKey] shk
		
		Knowledge:
		A: A,B,s;
		B: B,A,s;
		s: A,B,s;
		A,s share shk(A,s);
		B,s share shk(B,s)
		
		Actions:
		A -> B: A,NxNA
		B -> s: A,NxNA,B,NxNB
		s -> B: {|KxKAB,NxNB,A|}shk(B,s),{|NxNA,B,KxKAB|}shk(A,s)
		B -> A: {|NxNA,B,KxKAB|}shk(A,s),{|NxNA|}KxKAB,NxNB2
		A -> B: {|NxNB2|}KxKAB
		
		Goals:
		B authenticates s on KxKAB,A
		A authenticates s on KxKAB,B
		KxKAB secret between A,B,s
	\end{lstlisting}
	\caption{Carlsen protocol specification in \anb{} (concrete form)}
	\label{fig:anbx-example}
\end{figure}

The protocol involves three agents: two principals \(A\) and \(B\) who
wish to establish a fresh symmetric key \(KxKAB\), and a server \(s\)
that assists in the key distribution. The private function \(shk\),
declared with the \(\to^*\) annotation, allows to model a pre-shared symmetric
key between each agent and the server (\(shk(A,s)\) and \(shk(B,s)\)).
The protocol proceeds in five steps: \(A\) initiates with a nonce
\(NxNA\) sent to \(B\); \(B\) generates another nonce and forwards them to the server; the
server responds with two encrypted messages containing the new key
\(KxKAB\) and the nonces; \(B\) forwards the relevant part to \(A\); and
finally \(A\) acknowledges receipt using the new key.

The security goals declared in the specification are not directly
translated into defensive checks. Instead, the compiler derives the
checks from the protocol's type system and knowledge sets. These checks
are necessary, but not sufficient, consistency conditions: a captured
trace that fails a check is not a valid execution of the protocol, but
passing every check does not by itself establish that the protocol
achieves its declared security goals. The \anbx{} compiler translates
specifications such as the one in Figure~\ref{fig:anbx-example} into
executable Java code through a multi-stage pipeline, which we describe in
the following subsection.

\subsection{The \anbx{} Compiler}
\label{subsec:compiler-pipeline}

The \anbx{} compiler~\cite{anbx2015} translates high-level protocol
specifications into executable Java code through a pipeline consisting of multiple steps.
Our passive intruder reconstruction extends the pipeline with mechanisms
for generating intruder code and validating reconstructed serialised
objects.

\subsubsection{Compilation Stages}

The compilation consists of three main phases:

\begin{enumerate}
	\item \textit{\anbx{} to \anb{} Translation:} The abstract channel
	modes in the \anbx{} specification are compiled to concrete
	cryptographic constructions in the \anb{} language.
	
	\item \textit{\execnarr{} Generation:} The \anb{} protocol is
	translated to a typed \emph{executable narration} (\execnarr{}), an
	intermediate representation that captures the operational semantics
	of the protocol. The \execnarr{} includes defensive checks derived
	from the protocol's knowledge and type system, namely
	well-formedness, equality, inversion and seen checks.
	
	\item \textit{Java Code Generation:} The \execnarr{} is translated
	to Java code using template-based generation. The generated code
	uses the \anbxj{} security library~\cite{anbx2015} for cryptographic
	and communication operations.
\end{enumerate}

The defensive checks generated by the compiler are crucial for validating
incoming messages for honest agents, and captured traffic for the intruder:
\begin{itemize}
	\item \emph{Well-formedness checks} ensure that each received term
	has the expected type (e.g., a number in a numeric position, a symmetric key where a key is expected).
	\item \emph{Equality checks} verify that terms are equal as expected
	by the protocol (e.g., the nonce received from A matches the nonce
	sent by A).
	\item \emph{Inversion checks} verify that encrypted or signed
	messages can be correctly decrypted or verified using the appropriate
	keys.
	\item \emph{Seen checks} record the sequence numbers
	received so far and reject any that have been
	observed before, detecting replay when sequence numbers are used to
	implement freshness.
\end{itemize}

These checks instantiate the Dolev--Yao deduction rules presented in
Section~\ref{subsec:threat-model} on concrete messages, with well-formedness
checks imposing a stricter typing discipline than the type-free term
construction of the standard model. They form the basis of the validation
performed by the passive intruder.

\subsubsection{Passive Intruder Generation}

Two compilation modes support the reconstruction workflow. The first,
invoked with \texttt{-out:JavaDocker -jdockerpcap}, generates code for
protocol execution under a passive man-in-the-middle configuration within
a Dockerised environment, producing a PCAP file that captures all
protocol messages. The second, invoked with \texttt{-out:Java -objcheck},
generates the intruder code that reconstructs the captured trace from
extracted object files for each step. The reconstruction pipeline uses
the first mode to capture traffic and the second to reconstruct and
validate it; we describe both modes in detail in
Section~\ref{sec:methodology}.

The compiler also supports \texttt{-out:AnBxIntr}, which injects an
explicit intruder agent between every pair of communicating agents at the
\execnarr{} level (Section~\ref{subsec:intruder-injection}). This
transformation is applied internally by the reconstruction mode
(\texttt{-out:Java -objcheck}), which analyses the intruder-injected
narration to derive the receive actions the intruder must perform and the
defensive checks associated with them. The capture mode (\texttt{-out:JavaDocker -jdockerpcap}) does not rely on this transformation; the intruder is placed between the honest agents by the Docker network topology, described in Section~\ref{subsec:intruder-generation}.

\subsection{Java Serialisation}
\label{subsec:java-serialisation}

Our reconstruction pipeline relies on the ability to extract serialised Java objects from captured network traffic. Java serialisation is a mechanism for converting Java objects into a byte stream that can be transmitted over a network or persisted to disk~\cite{JavaSerialisation}; while the format is widely deployed, deserialising untrusted input has been the source of numerous remote-code-execution vulnerabilities~\cite{Sayar2023}.
A serialised stream begins with a four-byte header consisting of the magic
number \texttt{AC ED} and the protocol version \texttt{00 05}. The header
is followed by the serialised objects themselves; each object contributes
a class descriptor, its field data, and, where a value has already been
serialised earlier in the stream, a cross-reference to that earlier value
rather than a copy of it.

\begin{figure}[t]
	\centering
	\begin{tikzpicture}[
		font=\footnotesize\sffamily,
		box/.style={draw, rounded corners, align=left, inner sep=5pt},
		stream/.style={box, fill=blue!5, text width=6.2cm},
		table/.style={box, fill=green!5, text width=3.6cm},
		file/.style={box, fill=orange!5, text width=4.2cm},
		arrow/.style={->, >=stealth, shorten >=2pt, shorten <=2pt},
		]
		\node[stream] (stream) {
			\textbf{Captured serialised stream} \\[2pt]
			\texttt{AC ED 00 05} (stream header) \\
			\texttt{TC\_OBJECT} Object A \\
			\quad field \texttt{s}: \texttt{TC\_STRING} ``Alice'' (new handle H1) \\
			\quad \texttt{TC\_ENDBLOCKDATA} \\
			\texttt{TC\_OBJECT} Object B \\
			\quad field \texttt{s}: \texttt{TC\_REFERENCE} H1 (reuses H1) \\
			\quad \texttt{TC\_ENDBLOCKDATA}
		};
		
		\node[table, below=0.8cm of stream] (deser) {
			\textbf{Deserialiser} \\[2pt]
			Reads the whole stream \\[2pt]
			\textbf{Handle table} \\
			H1 $\rightarrow$ ``Alice'' \\
			\quad (first occurrence) \\[2pt]
			On \texttt{TC\_REFERENCE} H1, \\
			reuse the same object
		};
		
		\node[file, below=0.8cm of deser] (files) {
			\textbf{Per-step files} \\[2pt]
			\texttt{STEP\_1.ser}: Object A \\
			\quad \texttt{s} = ``Alice'' \\[2pt]
			\texttt{STEP\_2.ser}: Object B \\
			\quad \texttt{s} = ``Alice'' \\[2pt]
			{\scriptsize\sffamily\itshape No cross-file handle: each file is self-contained}
		};
		
		\draw[arrow] (stream) -- (deser)
		node[midway, right, font=\scriptsize\sffamily] {deserialise};
		\draw[arrow] (deser) -- (files)
		node[midway, right, font=\scriptsize\sffamily] {re-serialise};
	\end{tikzpicture}
	\caption{Java serialisation with \texttt{TC\_REFERENCE}. The first
		occurrence of a shared object is written in full and assigned a handle
		(here H1); later occurrences are written as \texttt{TC\_REFERENCE} H1.
		Reading the complete stream lets Java's deserialiser resolve H1
		automatically. When each top-level object is later written to its own
		step file, the shared object ("Alice") is materialised in each file, so the
		intruder role can read step files independently.}
	\label{fig:java-serialisation-tcreference}
\end{figure}
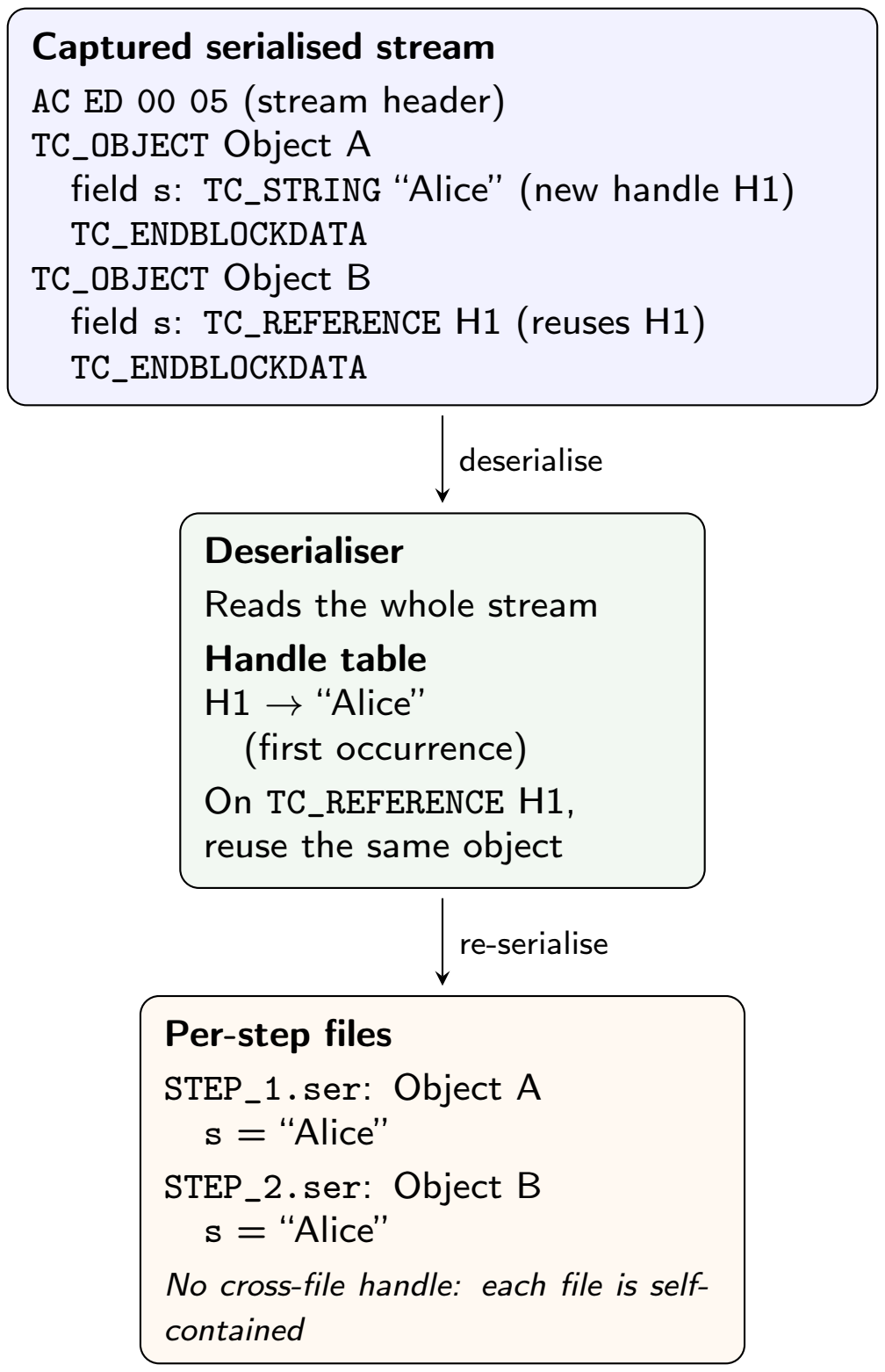

A key feature of Java serialisation is the handling of object graphs. If
an object contains references to other objects, the serialisation
protocol preserves the graph structure using \texttt{TC\_REFERENCE}
markers (\texttt{0x71}). When multiple objects are written to the same
stream, references from later objects can refer to earlier objects,
enabling efficient serialisation of complex data structures. For our
purposes, this means that a single TCP stream may contain multiple
serialised objects that reference each other. Figure
\ref{fig:java-serialisation-tcreference} illustrates the mechanism: the
first occurrence of a shared object is written in full and assigned a
handle; later occurrences are written as a \texttt{TC\_REFERENCE} to that
handle. Reading the complete stream lets Java's deserialiser resolve the
handle automatically, which is why our pipeline deserialises each
captured stream as a whole rather than attempting to split it into
objects by scanning for magic numbers.

The \texttt{ObjectInputStream} class handles deserialisation
automatically when used in a loop until \texttt{EOFException} is thrown.
However, Java serialisation introduces a subtle complication: when a
class overrides the \texttt{writeObject} method and enters
\emph{block-data mode}, the stream emits a \texttt{TC\_ENDBLOCKDATA}
(\texttt{0x78}) marker to delimit the block~\cite{JavaSerialisation}.
While these markers are correctly consumed during normal deserialisation,
they interact with the way Java's reader distinguishes between object and
primitive data when a captured stream is parsed offline; in the protocol
implementations evaluated here, this can produce
\texttt{OptionalDataException} during reconstruction. The pipeline must
therefore handle these markers explicitly, as we describe in
Section~\ref{sec:methodology}.

\paragraph{Example (multiple objects in one stream)}
Consider a stream carrying two serialised \texttt{AnBx\_Params} objects,
\(O_1\) and \(O_2\), written by a class that overrides
\texttt{writeObject}:

\[
\underbrace{\texttt{aced0005}\;\cdots\;\texttt{78}}_{O_1}\;
\underbrace{\texttt{73}\;\cdots\;\texttt{78}}_{O_2}
\]

Only \(O_1\) begins with the magic number; \(O_2\) starts with a
\texttt{TC\_OBJECT} marker (\texttt{0x73}) and no magic number. A reader
scanning for \texttt{aced0005} alone will therefore find exactly one
object per stream, regardless of how many were written, and cannot in
general recover the boundary between \(O_1\) and \(O_2\). A reader that
consumes the stream one object at a time handles the block-data markers
transparently and also maintains the reference table needed to resolve
\texttt{TC\_REFERENCE} markers such as those shown in
Figure~\ref{fig:java-serialisation-tcreference}. Because object
boundaries cannot be reliably identified from the stream alone, the
reconstruction pipeline defers object extraction to the \texttt{ObjCheck}
class, which reads each stream in a single pass and preserves the
reference table (Section~\ref{subsec:objcheck-mapping}).

\section{Methodology}
\label{sec:methodology}

This section presents our main contribution: an automated
pipeline that reconstructs passive Dolev--Yao intruder executions from
captured network traffic. We describe the complete workflow from PCAP to
executable Java code, focusing on the technical components that implement it:
PCAP-to-object extraction, connection pairing and session grouping,
candidate enumeration, channel-to-step matching, multi-candidate scoring and
selection, and passive intruder code generation.

\begin{figure}[ht]
	\centering
	\begin{tikzpicture}[
		node distance=1.5cm,
		every node/.style={rectangle, draw, minimum width=5.3cm, minimum height=0.55cm, align=center, rounded corners, font=\footnotesize\sffamily},
		arrow/.style={->, >=stealth, shorten >=2pt, shorten <=2pt},
		phase1/.style={fill=blue!10},
		phase2/.style={fill=green!10},
		phase3/.style={fill=orange!10},
		phase4/.style={fill=red!10},
		phase5/.style={fill=purple!10}
		]
		\node[phase1] (pcap) {PCAP file (\texttt{*\_DY.pcap})};
		\node[phase2, below of=pcap] (tshark) {PCAP-to-streams (\texttt{jpcap2ser.py} + \texttt{outparse.py})};
		\node[phase2, below of=tshark] (streams) {\texttt{*\_streams.json} + \texttt{STREAM\_X\_Y.ser}};
		\node[phase3, below of=streams] (sessionbuild) {Session build (\texttt{sessionbuild.py}):\\connections, sessions, candidates};
		\node[phase4, below of=sessionbuild] (passA) {\textbf{Pass A}: \texttt{ObjCheck} per candidate\\+ snapshot \texttt{*\_step\_snapshots/c\(i\)/}};
		\node[phase5, below of=passA] (passB) {\textbf{Pass B}: Java-gen + intruder\\per candidate \(\times\) session};
		\node[phase5, below of=passB] (select) {Score candidates, select winner};
		\node[phase5, right of=select, xshift=4.4cm] (execution) {Winner's reconstruction};
		
		\draw[arrow] (pcap) -- (tshark);
		\draw[arrow] (tshark) -- (streams);
		\draw[arrow] (streams) -- (sessionbuild);
		\draw[arrow] (sessionbuild) -- (passA);
		\draw[arrow] (passA) -- (passB);
		\draw[arrow] (passB) -- (select);
		\draw[arrow] (select) -- (execution);
		
		\node[draw=none, fill=none, anchor=west, xshift=0.2cm, yshift=0.3cm] at (pcap.north)          {\scriptsize Output of Stages 1--2};
		\node[draw=none, fill=none, anchor=west, xshift=0.2cm, yshift=0.3cm] at (tshark.north)        {\scriptsize Stage 3};
		\node[draw=none, fill=none, anchor=west, xshift=0.2cm, yshift=0.3cm] at (sessionbuild.north)  {\scriptsize Stage 4};
		\node[draw=none, fill=none, anchor=west, xshift=0.2cm, yshift=0.3cm] at (passA.north)         {\scriptsize Stage 5, Pass A};
		\node[draw=none, fill=none, anchor=west, xshift=0.2cm, yshift=0.3cm] at (passB.north)         {\scriptsize Stage 5, Pass B};
	\end{tikzpicture}
	\caption{Reconstruction pipeline from PCAP to a validated Java reconstruction.}
	\label{fig:pipeline-overview}
\end{figure}
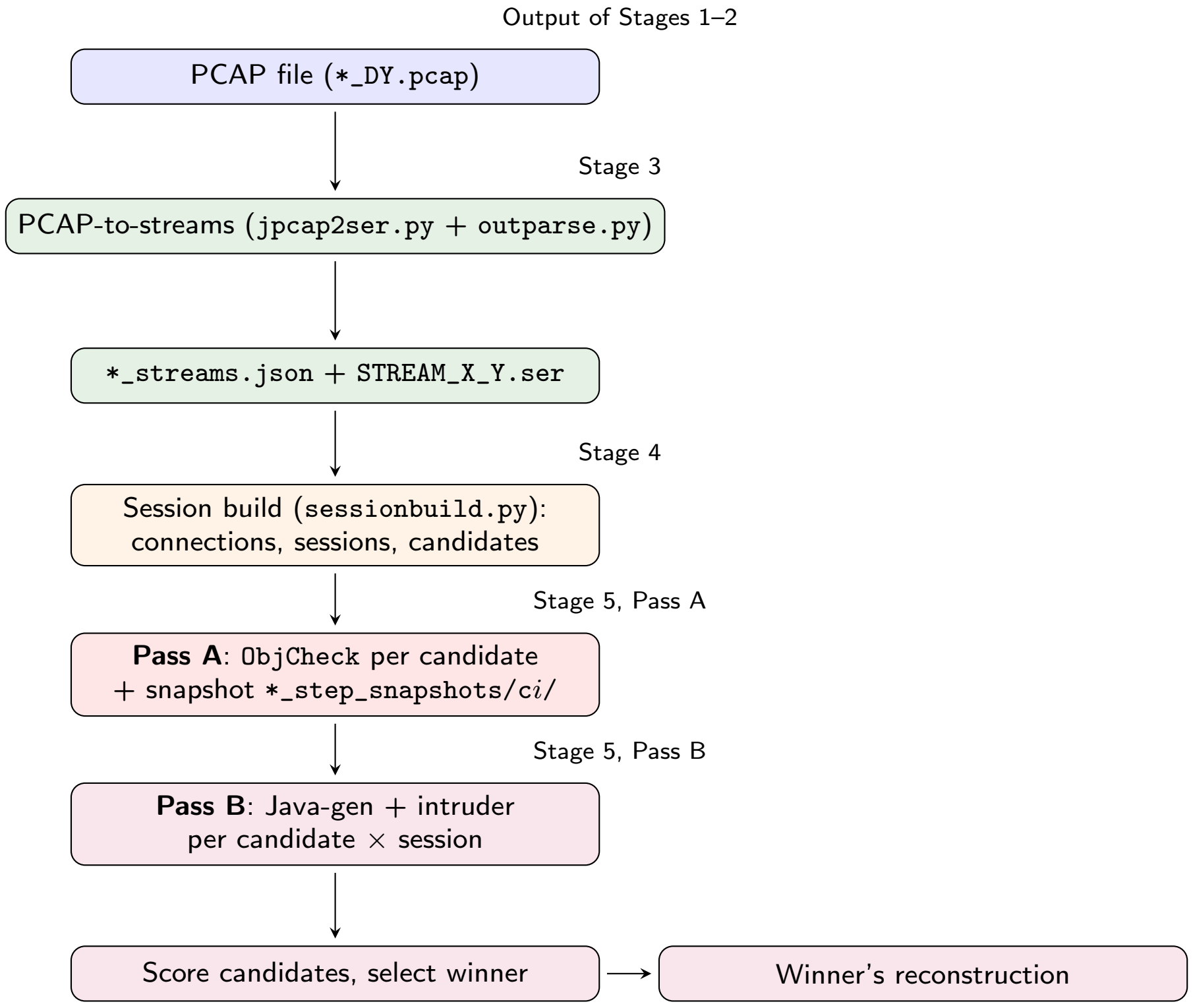

Figure~\ref{fig:pipeline-overview} illustrates the reconstruction methodology.
Stage~1 (\textit{JavaDocker generation}) and Stage~2 (\textit{Docker
	execution}) produce the PCAP. Stage~3 (\textit{PCAP-to-streams},
\texttt{jpcap2ser.py} + \texttt{outparse.py}) extracts per-TCP-stream
serialised object files together with per-stream metadata in
\texttt{\{Protocol\}\_streams.json}. Stage~4 (\textit{session build},
\texttt{sessionbuild.py}) pairs streams into connections, groups them into
sessions, and enumerates candidate channel assignments, writing one metadata
file per candidate. Stage~5 (\textit{candidate enumeration and selection},
orchestrated by \texttt{run\_test.py}) then iterates over the candidates
produced by Stage~4 in two passes: Pass~A runs the generated
\texttt{*\_ObjCheck} class for every candidate and snapshots its per-step
files; Pass~B runs the compiler and the passive intruder role for every
usable candidate, scores it, and selects the best-scoring arrangement.

\subsection{Notation}
\label{subsec:notation}

We write \(\mathcal{C}\) for a capture (the set of directional TCP streams
extracted from one PCAP), \(c\) for a bidirectional connection obtained by
pairing reverse streams, and \(\mathsf{sig}(c)\) for the signature of a
connection: its unordered endpoint IP pair together with the server port.
A protocol specification fixes a set of \emph{declared channels}, whose
distinct signatures determine \(K\), the number of connections per session.
A \emph{connected component} is a maximal set of connections linked
transitively by shared endpoint IPs. We write \(N\) for the number of sessions it
contains and \(M = N \cdot K\) for its number of connections. A
\emph{candidate arrangement} assigns to each of the \(N\) sessions one
connection for each of the \(K\) connection signatures; we write
\(\sigma_i\) for the \(i\)-th candidate arrangement, indexed from \(0\),
and refer to it as candidate~\(i\). Each candidate arrangement yields a set of per-session, per-step files and is scored by the lexicographic tuple described in Section~\ref{subsec:candidate-scoring}.

\subsection{Reconstruction Correctness}
\label{subsec:reconstruction-correctness}

The evaluation in Section~\ref{sec:results} uses the terms
\emph{complete reconstruction} and \emph{DY-consistent reconstruction}.
We define both precisely here.

\begin{definition}[Complete reconstruction]
	\label{def:complete}
	A candidate arrangement is \emph{complete} for a capture \(\mathcal{C}\) if every
	expected protocol step has exactly one recovered object associated with it.
\end{definition}

\begin{definition}[DY-consistent reconstruction]
	\label{def:dy-consistent}
	A candidate arrangement is \emph{DY-consistent} if every
	compiler-generated passive-intruder check executed on the recovered
	objects passes.
\end{definition}

\begin{definition}[Successful reconstruction]
	\label{def:successful}
	A capture \(\mathcal{C}\) is \emph{successfully reconstructed} if the selected
	candidate arrangement is both complete and DY-consistent.
\end{definition}

Definitions~\ref{def:complete}--\ref{def:successful} are consistency
conditions on the captured trace, not a proof of security. A trace that
satisfies them is consistent with the specification's type system,
equality constraints, and freshness requirements; a trace that fails one
of them is not a valid execution of the protocol under the specification.

\subsection{Extraction of Serialised Objects}
\label{subsec:object-extraction}

The extraction stage processes a PCAP file to produce serialised Java objects.
It is implemented by two Python scripts: \texttt{jpcap2ser.py} orchestrates
the extraction, while \texttt{outparse.py} performs the parsing and emits
metadata.

The \texttt{jpcap2ser.py} script invokes TShark~\cite{tshark} to extract TCP
stream data:

\begin{lstlisting}[basicstyle=\footnotesize\ttfamily,breaklines=true]
	tshark -r input.pcap -Y "tcp.payload" -2 -T fields \
	-e tcp.stream -e ip.addr -e data \
	-e ip.src -e ip.dst -e tcp.srcport -e tcp.dstport \
	-e frame.time_epoch
\end{lstlisting}

The \texttt{-Y "tcp.payload"} filter selects packets that carry
application-layer data, and the \texttt{-e data} flag extracts that
application-layer payload itself. The \texttt{-2} flag enables two-pass
analysis, which the pipeline uses to ensure TCP segments are reassembled
consistently before the \texttt{data} field is extracted. This command
extracts:

\begin{itemize}
	\item the TCP stream index (\texttt{tcp.stream}), which groups packets
	belonging to the same TCP connection;
	\item the IP addresses involved (\texttt{ip.addr}), used to identify the
	communicating agents;
	\item the raw application data (\texttt{data}), which contains the
	serialised Java objects;
	\item the per-packet source and destination IPs and ports
	(\texttt{ip.src}, \texttt{ip.dst}, \texttt{tcp.srcport},
	\texttt{tcp.dstport}), used to pair bidirectional streams into
	connections and to compute channel signatures
	(Section~\ref{subsec:session-grouping});
	\item the packet timestamp (\texttt{frame.time\_epoch}), used for
	tie-breaking during connection pairing and as a per-stream start/end
	timestamp.
\end{itemize}

We extract the \texttt{data} field rather than \texttt{tcp.payload} because \texttt{data} is
populated by Wireshark's fallback dissector for payloads that no other
dissector recognises. It therefore yields the raw application-layer bytes
without any upper-layer protocol headers that a dissector might otherwise
consume. Extracting \texttt{tcp.payload} directly can include such headers when
the beginning of the payload happens to be recognised as another protocol,
which interferes with object detection.

The \texttt{outparse.py} script processes the TShark output, grouping packets
by TCP stream index and saving each complete directional TCP stream as a
\texttt{STREAM\_X\_Y.ser} file, where \texttt{X} is the one-based TCP stream
number (TShark reports \texttt{tcp.stream} as zero-based, so
\texttt{tcp.stream=0} becomes \texttt{STREAM\_1\_Y.ser}) and \texttt{Y} is
the direction within the stream. The pipeline always invokes
\texttt{outparse.py} with \texttt{--allstreams}, so every directional stream
in the PCAP is saved. As a heuristic, \texttt{outparse.py} also checks
whether the application-layer payload of each packet begins with the Java
serialisation magic number (\texttt{aced0005}); this is used only to
characterise each stream, not to delimit objects within it.

When invoked with \texttt{----emit-meta}, \texttt{outparse.py} additionally
emits \texttt{\{Protocol\}\_streams.json}, one record per directional TCP
stream, containing the stream filename, the TCP stream index, the source and
destination IPs and ports, the first and last observed timestamps, and the
number of packets whose application-layer payload begins with a Java
serialisation magic number. This metadata is the input to the session build
stage.

A key challenge in this stage is that a single directional stream may contain
multiple serialised objects. When more than one object is written to the same
\texttt{ObjectOutputStream}, only the first begins with the magic number, and
no explicit markers separate the objects thereafter. Rather than attempting
to delimit objects at the byte level, \texttt{outparse.py} saves each stream
in full and defers object extraction to the \texttt{ObjCheck} class, which
reads the stream once and preserves the Java serialisation reference table
(see Section~\ref{subsec:objcheck-mapping}).

\paragraph{Example}
For the Carlsen capture at \(N=2\), each session contributes four
directional streams, eight in total. In each session, the \(A \to B\)
direction of the \(A \leftrightarrow B\) connection carries two objects
(steps 1 and 5); the other three streams of the session carry one object
each. This multi-object stream is the case that motivates deferring object
extraction to \texttt{ObjCheck}: the second object does not begin with the
magic number and cannot be delimited by scanning alone.

\subsection{Connection Pairing and Session Grouping}
\label{subsec:session-grouping}

Stage~4, implemented by \texttt{sessionbuild.py}, consumes
\texttt{\{Protocol\}\_streams.json} together with the protocol's
\texttt{.properties} file and groups the observed streams of the capture
\(\mathcal{C}\) into protocol sessions. The \texttt{.properties} file is the concrete representation of the
deployment metadata introduced in Section~\ref{subsec:threat-model}: it
records the mapping from IP addresses and ports to protocol roles, the
channel structure, and the expected channel order. In our implementation
it is generated by the \anbx{} compiler, but its content is
deployment-specific rather than compiler-specific, and the same
information would be available to an analyst from network metadata,
protocol documentation, or prior observation. From these definitions
\texttt{sessionbuild.py} computes \(K\), the number of distinct connection
signatures that constitute one protocol session. A connection signature is the
unordered pair of endpoint IPs together with the server port; channels that
share the same signature are counted once, so \(K\) is not in general equal to
the number of declared channels. \(K\) is therefore a per-session quantity: a
capture containing \(N\) sessions comprises \(N \cdot K\) connections in total.

\paragraph{Connection pairing}
Directional TCP streams are first paired into \emph{connections}. Each
stream has an oriented endpoint tuple
\((\mathit{src\_ip},\mathit{src\_port},\mathit{dst\_ip},\mathit{dst\_port})\);
a connection is keyed by the unordered pair of its endpoints, so the two
directional streams that make up a bidirectional connection share a key
even though their tuples are exact reverses of each other.
If more than one reverse candidate exists, the peer whose per-stream
\texttt{t\_first} timestamp is closest to the stream's \texttt{t\_first} is
selected. The resulting connection \(c\) records its two directional stream
files, its endpoints, its signature \(\mathsf{sig}(c)\), and its start and
end timestamps, denoted \texttt{t\_start} and \texttt{t\_end}. The
connection's \texttt{t\_start} is the smaller of its two streams'
\texttt{t\_first} values. The direction of the stream whose \texttt{t\_first} is
smaller is labelled \emph{ab}; the reverse direction is \emph{ba}.

\paragraph{Session grouping}
Before grouping, connections whose signature does not match any declared
channel are discarded. This filters out traffic that happens to share
endpoints with the protocol but is not part of it, such as the Docker network
handshake, health checks, or unrelated background traffic.

The remaining connections are then grouped using \emph{connected components}
over shared agent IPs: two connections belong to the same connected component if they
share at least one endpoint IP. A connected component is therefore a set of
connections transitively linked through shared endpoints, not a set of
agents or sessions.  
In the typical capture, where a fixed set of agent IPs participates in the
protocol execution, this yields a single connected component containing all the
connections of that execution, across all of its parallel sessions; the
connected component therefore spans every agent role in the protocol, not a single
agent.  

For a well-formed single-session trace the connected component contains exactly \(K\)
connections. When the PCAP contains \(N\) parallel sessions, a connected component may
contain \(M = N \cdot K\) connections, all sharing the same agent identities
and using a distinct ephemeral port per session. The number of connections in
a connected component relative to \(K\) distinguishes three cases:

\begin{itemize}
	\item \(M = K\): the connected component is a single session. It is emitted
	directly, tagged with confidence level \texttt{high}.
	\item \(M > K\): the connected component contains \(N = M/K\) interleaved
	sessions. It is split by enumerating candidate arrangements, as
	described in Section~\ref{subsec:candidate-enumeration}. The resulting
	sessions are tagged with confidence level \texttt{medium} and a reason
	such as \texttt{signature-split (\(m\) arrangements)}.
	\item \(M < K\): the connected component contains fewer connections than a single
	session requires. A warning is emitted; in \texttt{--strict} mode the
	protocol is aborted.
\end{itemize}

The grouping is deliberately conservative: the time-based
\texttt{t\_start} ordering (the connection-level timestamp defined above)
is not trusted as a correctness criterion. It is used only as a
tie-breaker when pairing reverse streams, and to sort connections within each
signature before candidate enumeration. The final choice of arrangement is
made later, by scoring the candidate reconstructions
(Section~\ref{subsec:candidate-scoring}).

\begin{figure}[ht]
	\centering
	\begin{tikzpicture}[
		agent/.style={circle, draw, thick, minimum size=0.7cm, inner sep=1pt, font=\small},
		link/.style={draw, thick, shorten >=4pt, shorten <=4pt, <->},
		]
		\node[agent] (A) at (0,0) {A};
		\node[agent] (B) at (3,0) {B};
		\node[agent] (s) at (6,0) {s};
		
		\draw[link, red,  bend left=20]  (A) to node[above, font=\scriptsize]{$c_1$} (B);
		\draw[link, blue, bend right=20] (A) to node[below, font=\scriptsize]{$c_2$} (B);
		\draw[link, red,  bend left=20]  (B) to node[above, font=\scriptsize]{$c_3$} (s);
		\draw[link, blue, bend right=20] (B) to node[below, font=\scriptsize]{$c_4$} (s);
		
		\draw[dashed, thick, rounded corners] (-0.7,-1.0) rectangle (6.7,1.0);
		\node[anchor=south east, font=\scriptsize] at (6.6,0.9) {connected component};
		\node[font=\scriptsize, red]  at (1.5,-0.8) {session 1};
		\node[font=\scriptsize, blue] at (4.5,-0.8) {session 2};
	\end{tikzpicture}
\caption{The Carlsen connected component at \(N = 2\). The colour
	grouping is candidate~0, the identity arrangement; the pipeline
	also evaluates permutations of this assignment
	(Section~\ref{subsec:candidate-scoring}).}
	\label{fig:component-example}
\end{figure}
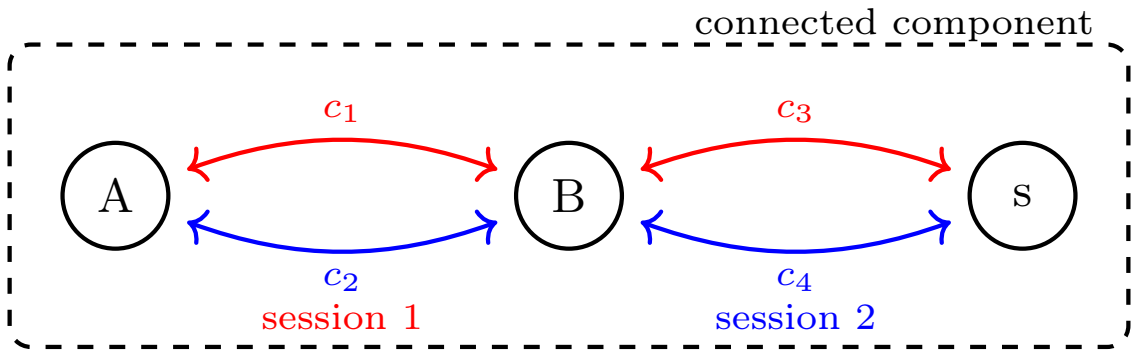

\paragraph{Example} Figure~\ref{fig:component-example} shows the shape of a typical connected
component for the Carlsen protocol at \(N = 2\). Carlsen declares two
connection signatures (\(K = 2\)), the A\(\leftrightarrow\)B and
B\(\leftrightarrow\)s connections, so the component contains
\(M = N \cdot K = 4\) connections, all transitively linked through B.
The pipeline must recover which A\(\leftrightarrow\)B connection belongs to
which B\(\leftrightarrow\)s session; this is the step that candidate
enumeration and scoring perform.

\subsection{Candidate Enumeration and Channel Assignment}
\label{subsec:candidate-enumeration}

A connected component that contains exactly \(K\) connections is a single
session and is emitted directly (Section~\ref{subsec:session-grouping}).
When it contains more connections than that, the pipeline must decide how
they divide into sessions. Let \(M\) be the number of connections in the
component. Since each session contributes one connection per signature,
the number of sessions is \(N = M/K\), where \(K\) is the number of
connection signatures per session. 
A component whose size is not a multiple of \(K\) is discarded with a
warning, since the capture is then inconsistent with the protocol's
declared channel structure.

The pipeline groups the connections by
signature, so that each of the \(K\) signatures holds exactly \(N\)
connections, and sorts the connections within each signature by
\texttt{t\_start}.

For \(K = 1\) there is a single arrangement, since \((N!)^0 = 1\) and there
is nothing to permute: with only one signature, each of the \(N\) sessions
receives a distinct connection and the assignment is forced. For \(K > 1\),
one signature is fixed as the reference. Its \(N\) connections, already
sorted by \texttt{t\_start}, are numbered \(1\) to \(N\), and connection
\(i\) in this ordering defines session \(i\). The remaining \(K-1\)
signatures are then assigned to sessions by permutation: for each such
signature, any of the \(N!\) bijections from its \(N\) connections to the
\(N\) sessions is admissible. The total number of candidate arrangements is
therefore
\[
(N!)^{K-1}.
\]
Each arrangement thus assigns, for each of the \(N\) sessions, one
connection for each of the \(K\) signatures.

Because the candidate space grows factorially, \texttt{sessionbuild.py}
applies a cap \texttt{MAX\_ARRANGEMENTS}. When the raw number of arrangements
\((N!)^{K-1}\) is at most the cap, the full space is enumerated. Otherwise the pipeline falls back to a heuristic subset of arrangements; this path is not exercised by our evaluation.
The cap is set to \(120\), which keeps full enumeration feasible
up to \(N \le 5\) for \(K = 2\), \(N \le 3\) for \(K = 3\), and \(N \le 2\)
for \(K \ge 4\). For the Carlsen-style protocols used in our evaluation
(\(K = 2\)), this corresponds to \(N \le 5\); the raw space for \(N = 10\)
would be \(10! = 3{,}628{,}800\) arrangements, which is infeasible at
Stage~4. Larger \(N\) requires content-based pruning and is left as future
work (Section~\ref{sec:conclusion}).

Each candidate arrangement is written to a flat TXT file
\texttt{\{Protocol\}\_session\_meta\_c\(i\).txt}; collectively, these are the
files consumed by Stage~5. Each file identifies the protocol, the input PCAP,
the number of sessions, and the candidate index \(i\), and then, for every
session, lists the connections that the arrangement assigns to it, one line
per channel, of the form
\[
\texttt{channel=<channel\_name>,<conn\_id>,<dir>,<stream\_file>},
\]
where \texttt{conn\_id} is the identifier of the connection, \texttt{dir} is
its direction (\texttt{ab} or \texttt{ba}), and \texttt{stream\_file} is the
directional stream file from which the objects for that channel are to be
read by the object checker.

\paragraph{Example} The Carlsen component of
Figure~\ref{fig:component-example} has \(K = 2\) and, at \(N = 2\), two
connections per signature. Denote by \(c_1, c_2\) the A\(\leftrightarrow\)B
connections in \(\mathsf{t_{start}}\) order, and by \(c_3, c_4\) the
B\(\leftrightarrow\)s connections. The candidate arrangements are
\[
\sigma_0 : \;\; \text{session 1} = \{c_1, c_3\},\;\;
\text{session 2} = \{c_2, c_4\},
\]
\[
\sigma_1 : \;\; \text{session 1} = \{c_1, c_4\},\;\;
\text{session 2} = \{c_2, c_3\}.
\]
The colour grouping in Figure~\ref{fig:component-example} corresponds to
\(\sigma_0\), the identity arrangement, in which each signature's
connections are paired in their \texttt{t\_start} order.
The pipeline evaluates both arrangements and selects the one with the
higher score under the tuple of Section~\ref{subsec:candidate-scoring}.

\subsection{ObjCheck and Channel-to-Step Mapping}
\label{subsec:objcheck-mapping}

In our reconstruction pipeline, each protocol step corresponds to exactly one
serialised Java object that is sent or received. We therefore use the terms
\emph{step} and \emph{object} interchangeably when referring to the number of
messages in a protocol execution.

A critical challenge is determining the correct protocol step for each
extracted object. The order in which streams appear in the PCAP does not
necessarily correspond to the protocol's step order, and a single
directional stream may carry objects belonging to different steps. The \texttt{ObjCheck} class, generated by the \anbx{} compiler, resolves this mapping using the candidate-specific metadata file and the
compiler-generated step-to-channel map. The compiler generates the \texttt{Carlsen\_Channels} enum from the channel definitions in the \texttt{.properties} file, preserving their order. This shared ordering is what allows the compiled code (which uses the enum), the \texttt{.properties} file (read by the extraction stage), and the captured streams to agree on channel identifiers. For the Carlsen
protocol, the \texttt{ObjCheck} step-to-channel map is:

\begin{lstlisting}[language=Java,basicstyle=\footnotesize\ttfamily,breaklines=true]
	static Map<Carlsen_Steps, Carlsen_Channels> channelStepMap = new HashMap<>();
	static {
		channelStepMap.put(Carlsen_Steps.STEP_1, Carlsen_Channels.ROLE_B_channel_ROLE_A_Server_Insecure);
		channelStepMap.put(Carlsen_Steps.STEP_2, Carlsen_Channels.ROLE_s_channel_ROLE_B_Server_Insecure);
		channelStepMap.put(Carlsen_Steps.STEP_3, Carlsen_Channels.ROLE_B_channel_ROLE_s_Client_Insecure);
		channelStepMap.put(Carlsen_Steps.STEP_4, Carlsen_Channels.ROLE_A_channel_ROLE_B_Client_Insecure);
		channelStepMap.put(Carlsen_Steps.STEP_5, Carlsen_Channels.ROLE_B_channel_ROLE_A_Server_Insecure);
	}
\end{lstlisting}

\texttt{ObjCheck} also constructs the inverse map
\texttt{channelStepMapInverse()}, which associates each channel with the
ordered list of protocol steps expected on that channel. For each channel
entry in the candidate's metadata file, \texttt{ObjCheck} opens the
corresponding stream file with a single \texttt{ObjectInputStream} and
assigns the deserialised objects to consecutive steps for that channel, in
ascending step order as recorded by the inverse map. If the number of objects
and the number of expected steps differ, a mismatch is reported and the
assignment is truncated to the minimum of the two. The missing steps are later reflected in the candidate's score
(Section~\ref{subsec:candidate-scoring}).
The result is a set of per-session, per-step files
\texttt{\{Protocol\}\_<session>\_STEP\_<n>.ser}. These are the files consumed
by the generated passive intruder role. Multi-object streams are handled in the same pass, since the inverse map records the ordered list of steps expected on each channel.

\paragraph{Block-data handling}
A further challenge is Java serialisation's \texttt{TC\_ENDBLOCKDATA}
markers (\texttt{0x78}), which delimit custom \texttt{writeObject()}
implementations (Section~\ref{subsec:java-serialisation}). In the protocol
implementations evaluated here, custom serialisation and block-data mode
can produce \texttt{OptionalDataException} during offline extraction when
object boundaries are reconstructed from captured streams. The \texttt{readObjectsFromFile} method in \texttt{ObjCheck} handles
\texttt{OptionalDataException} by skipping the indicated number of bytes.
In principle, one could parse the serialised stream manually and adjust the
\texttt{TC\_REFERENCE} markers to point to the correct objects; however, this
would require low-level byte-level parsing and a global analysis of the object
graph across the entire captured traffic, as references are relative to the
position of objects within the stream (see Figure \ref{fig:java-serialisation-tcreference}). Our approach avoids this complexity by
leveraging Java's built-in deserialisation mechanism on complete streams,
which automatically resolves references without manual intervention.

When each object is later saved as a separate
\texttt{\{Protocol\}\_<session>\_STEP\_<n>.ser} file, references between
objects are not preserved across step files. This is acceptable because the
intruder role reads each step file independently and does not rely on object
identity being preserved across step files. For parallel sessions, the
situation is even simpler: each session uses a distinct TCP connection and
therefore a distinct stream file, so objects from different sessions never
shared a reference table in the first place.

\paragraph{Separation of ObjCheck from code generation}
One implementation detail is important for the orchestration. When the
compiler is invoked with \texttt{-out:Java -objcheck}, it may overwrite files
such as \texttt{*\_Channels.java}. Running a later \texttt{ObjCheck} after
that would therefore be unsafe. The pipeline consequently separates all object
checks from all Java generation and execution; the two passes are described in
the following subsection. 

\paragraph{Example}
For the Carlsen session in Figure~\ref{fig:component-example}, the inverse
map associates the A\(\leftrightarrow\)B channel with the ordered pair
\texttt{[STEP\_1, STEP\_5]}, and each of the other three channels with a
single step. The candidate metadata in turn specifies one stream file per
channel. Applying the two tables together, \texttt{ObjCheck} opens each
stream file once and writes the objects it reads to the corresponding step
files: the A\(\leftrightarrow\)B stream yields two step files
(\texttt{STEP\_1} and \texttt{STEP\_5}), and each of the other three streams
yields one. The result is five step files per session, one for each
protocol step.

\subsection{Multi-Candidate Scoring and Selection}
\label{subsec:candidate-scoring}

Candidate selection is orchestrated by \texttt{run\_test.py}
in two passes over the candidate arrangements produced by \texttt{sessionbuild.py}.

\paragraph{Pass A: ObjCheck per candidate}
For each candidate index \(i\), the pipeline copies
\texttt{\{Protocol\}\_session\_meta\_c\(i\).txt} to the working path
\texttt{\{Protocol\}\_session\_meta.txt}, removes any existing
\texttt{\{Protocol\}\_*\_STEP\_*.ser} files, and runs \texttt{ant objcheck}, 
which produces the per-step \texttt{.ser} files for that candidate.
If the candidate TXT is missing, or if \texttt{ObjCheck} produces no
\texttt{STEP} files, the candidate is recorded with a sentinel score
\((-1,0,0)\) and skipped. Otherwise, the produced \texttt{STEP} files are
copied to a snapshot directory
\texttt{\{Protocol\}\_step\_snapshots/c\(i\)/}. The snapshot is essential: it
preserves each candidate's object assignment independently of later compiler
invocations that would otherwise overwrite \texttt{*\_Channels.java}.

\paragraph{Pass B: Java generation and passive intruder per candidate}
For each candidate that survived Pass~A, the pipeline restores the snapshot,
copies the candidate TXT back to the working path, and then iterates over the
sessions. For each session it runs

\begin{lstlisting}[basicstyle=\footnotesize\ttfamily,breaklines=true]
	anbxc <Proto>.AnBx -out:Java -objcheck -nocfgmsg -silent \
	-securityProvider <BC|default> -objchecksession <sid>
\end{lstlisting}

where \texttt{BC} selects the Bouncy Castle provider and \texttt{default}
selects the standard JDK provider, followed by

\begin{lstlisting}[basicstyle=\footnotesize\ttfamily,breaklines=true]
	ant -buildfile <src>/build.xml runagents
\end{lstlisting}

The captured standard output and standard error are retained for scoring.

\paragraph{Scoring}
The passive intruder emits check results using the markers
\texttt{CHECK \# <id>} and \texttt{check FAILED!}. These markers are emitted
by the check functions in the \anbxj{} library and parsed by the orchestrator;
they form a light-weight internal interface between the generated code and
the scoring logic, and can be replaced with a structured report if the
toolchain is extended. For each candidate, the pipeline counts, per session,
the total number of checks and the number of failed checks. A session with no
recorded checks is penalised more heavily than one with failing checks: it
decrements the clean-session count rather than merely failing to increment
it. The candidate score is the lexicographic tuple
\[
(\mathit{n\_clean\_sessions},\; -\mathit{total\_failed\_checks},\;
\mathit{total\_checks}),
\]
where \(\mathit{n\_clean\_sessions}\) is the number of sessions that ran at
least one check and failed none, minus the number of sessions that ran no
check at all. Higher tuples are better. The selected candidate is the one with
the maximum score; ties are broken by the lowest candidate index.

Object recovery enters the score indirectly. A missing object means the
checks that would have been exercised on it are not counted, which lowers
\(\mathit{total\_checks}\); if it leaves a session with no checks at all,
\(\mathit{n\_clean\_sessions}\) is decremented. A candidate that recovers
fewer objects therefore scores lower, all else being equal. The score
function captures the intuition that a correct arrangement should reconstruct
every session and pass every defensive check: the negative total-failed term
penalises incorrect pairings, and the total-checks term rewards arrangements
that produce more reconstruction evidence.

Scoring is performed for every candidate, not only the winner, so that the
alternatives can be compared on the same capture. The procedure selects the
correct arrangement whenever it is among the enumerated candidates and is
the unique maximiser of the score tuple; it does not guarantee correctness
when the correct candidate is absent or when ties occur, in which case the
lowest-index candidate is selected.

\paragraph{Winner selection}
After selection, the pipeline logs the winner's channel assignment (one line
per channel per session), overrides the aggregate return codes with the
winner's per-stage return codes, and copies the winning candidate TXT to the
final \texttt{\{Protocol\}\_session\_meta.txt}. This ensures that the CSV row
and the logged reconstruction summary describe the candidate that was actually
shipped, rather than the last candidate visited.

The multi-candidate approach is what makes the pipeline robust to the
failures of the time-based heuristic used during pairing. The
\texttt{t\_first} timestamps of the per-session connections can differ by
only a few microseconds, so the identity arrangement (which places each
signature's connections in the order of their \texttt{t\_start}
timestamps) is not guaranteed to be the correct one, and the winning
candidate is not fixed across session counts of the same protocol. The
scoring loop is therefore not an optimisation on top of the heuristic but
the mechanism that allows the reconstruction to correct the heuristic
whenever the correct arrangement is among the enumerated candidates. What
the score table contributes is not a fixed index but the identification,
for each individual reconstruction, of the best-scoring arrangement under
the scoring tuple defined above.

\paragraph{Example} For the Carlsen component of
Figure~\ref{fig:component-example} at \(N = 2\), candidate~0 (the identity
arrangement) scores \((0, -2, 26)\) and candidate~1 scores \((2, 0, 26)\).
Both candidates execute the same 26 checks, but candidate~1 leaves both
sessions clean whereas candidate~0 leaves two failed checks; since the
comparison is lexicographic, candidate~1 is selected. This is the 
arrangement \(\sigma_1\) of the enumeration example 
in Section~\ref{subsec:candidate-enumeration}.

\subsection{Intruder Code Generation}
\label{subsec:intruder-generation}

Pass~B of Stage~5 invokes the compiler for every candidate and executes the
generated passive intruder role on the candidate's step files; the checks the
role performs feed the scoring. This subsection explains how the compiler
transforms the protocol specification into that role, the
\texttt{ROLE\_intr} class.

\subsubsection{Intruder Injection}
\label{subsec:intruder-injection}

The compiler provides a transformation, selected by \texttt{-out:AnBxIntr},
that injects an intruder agent between every pair of communicating agents at
the \execnarr{} level. The reconstruction mode (\texttt{objcheck}) applies
this transformation internally as an intermediate step, so the transformed
narration is what the compiler analyses when generating the passive intruder
role's checks; the capture mode relies instead on the Docker network
topology, not on this transformation. Each send action in the original
protocol is split into two actions:

\begin{enumerate}
	\item \texttt{A:send(intr,M)} --- the original sender sends to the intruder;
	\item \texttt{intr:send(B,M)} --- the intruder forwards to the intended
	recipient.
\end{enumerate}

Crucially, the intruder is treated as just another agent role in the
transformed protocol. The compiler's defensive check generation applies
uniformly to every agent: for each \texttt{receive()} action, it generates
well-formedness checks (\texttt{wff}), equality checks (\texttt{eq}),
inversion checks (\texttt{inv}) and \texttt{seen} checks on sequence numbers,
based on the types of the received terms, the expected values, and the
cryptographic operations required. Thus, the intruder
automatically receives the defensive checks derived by the
compiler from the specification, matching honest agents' behaviour.

For the Carlsen protocol, the original action:

\begin{lstlisting}[basicstyle=\footnotesize\ttfamily,breaklines=true]
	1    | A: send(B;<A,NxNA>)
	1    | B: VAR_B_R1 := receive()
	1.1  | B: wff(proj[2/2][VAR_B_R1])
	1.2  | B: eq(A,proj[1/2][VAR_B_R1])
\end{lstlisting}

becomes:

\begin{lstlisting}[basicstyle=\footnotesize\ttfamily,breaklines=true]
	1    | A: send(intr;<A,NxNA>)
	1    | intr: VAR_INTR_R1 := receive()
	1.1  | intr: wff(proj[2/2][VAR_INTR_R1])
	1.2  | intr: eq(A,proj[1/2][VAR_INTR_R1])
	2    | intr: send(B;<A,proj[2/2][VAR_INTR_R1]>)
	2    | B: VAR_B_R2 := receive()
	2.1  | B: wff(proj[2/2][VAR_B_R2])
	2.2  | B: eq(A,proj[1/2][VAR_B_R2])
\end{lstlisting}

The intruder's checks at steps~1.1 and~1.2 are generated automatically by the
compiler's standard check derivation mechanism, not by copying from the honest
agent's checks at steps~2.1 and~2.2. Both sets of checks are derived from the
same underlying protocol model and type information, ensuring consistency
across all agents.

\paragraph{Correspondence with Dolev--Yao analysis}
The check at step~1.2 is the compiled form of a constraint derived from
the Dolev--Yao analysis of the specification. The projection
\texttt{proj[1/2][VAR\_INTR\_R1]} extracts the first component of the
observed tuple, matching the Dolev--Yao projection rule on the received
message. The equality check on the extracted value records the
constraint the specification places on this component: it must carry the
identity of agent~A. Each check the compiler generates corresponds to one
such constraint, and the check's outcome records whether the recovered
observation satisfies it. This is the sense in which the compiled intruder mirrors a Dolev--Yao
intruder whose deduction has been performed in advance: the deduction
rules (projection, decryption) are evaluated on the recovered message,
and the check functions record the constraints under which the deduced
values are consistent with the specification.

\subsubsection{Capture and Reconstruction Modes}

The capture mode, invoked with \texttt{-out:JavaDocker -jdockerpcap},
generates Java code that executes the protocol within Docker containers. The
Docker network is arranged as a star whose centre is the \texttt{DY}
container, through which all traffic between the honest agents passes; this
container runs a packet capture and records the traffic to a PCAP file. 
The reconstruction
mode, invoked with \texttt{-out:Java -objcheck}, generates the
\texttt{ROLE\_intr} class, which reads
\texttt{\{Protocol\}\_<session>\_STEP\_<n>.ser} files and reconstructs the
captured trace offline.

When generating code with \texttt{-objcheck}, the compiler applies a further transformation to the intruder-injected \execnarr{}. First, all \texttt{send()} actions are removed. 
Second, only the
intruder's receive actions are retained. In the transformed narration these
fall on odd-numbered positions (step~1: \(A \to intr\); step~3:
\(B \to intr\); step~5: \(s \to intr\); and so on), because each original
send becomes a pair of actions (sender-to-intruder, then
intruder-to-recipient) and the intruder's receive is the first of the pair.
Third, the retained steps are renumbered consecutively: odd steps \(s\)
become \((s+1)/2\), preserving the order of observation.

For the Carlsen protocol, the original intruder receive steps
\((1,3,5,7,9)\) become \texttt{STEP\_1} through \texttt{STEP\_5} in the
generated code. When the reconstruction is run across \(N\) parallel
sessions, the compiler is invoked once per session with
\texttt{-objchecksession <sid>}, and the generated \texttt{ROLE\_intr}
processes the corresponding session's step files in the working directory.

The compiler also determines the expected type of each step's payload by
analysing the expression in the corresponding \texttt{send()} action. If the
expression is a signed message (\texttt{NESign}), the step expects a
\texttt{Crypto\_SignedObject}; if it is encrypted (\texttt{NEEnc} or
\texttt{NEEncS}), the step expects a \texttt{Crypto\_SealedObject};
otherwise, it expects an \texttt{AnBx\_Params} (a tuple of values). This type
information is used in the generated \texttt{executeStep} method to cast
deserialised objects correctly, as shown in the code listing below:

\begin{lstlisting}[language=Java,basicstyle=\footnotesize\ttfamily,breaklines=true]
case STEP_1:
	AnBx_Params VAR_STEP_1 = (AnBx_Params)
	AnB_Session.readObject(filename);
	break;
case STEP_5:
	Crypto_SealedObject VAR_STEP_5 = (Crypto_SealedObject)
	AnB_Session.readObject(filename);
	break;
\end{lstlisting}

\subsubsection{Generated Code}
\label{subsec:generated-code}

The \texttt{ROLE\_intr} class (Figure~\ref{fig:intr-code}) reads
serialised objects from the per-step files and performs the defensive
checks derived by the compiler from the specification. The intruder does
not send messages to the network; it merely validates the captured trace.
It is important to distinguish this validation from structural
deserialisation. Deserialisation ensures that the bytes are well-formed
Java objects; the defensive checks verify that these objects conform to
the protocol's type system, equality constraints, and freshness
requirements. The latter is what determines whether a captured trace
constitutes a valid protocol execution.

\begin{figure}[ht]
	\begin{lstlisting}[language=Java,basicstyle=\footnotesize\ttfamily,breaklines=true,tabsize=2]
public final class Carlsen_ROLE_intr extends AnB_Protocol<...> {
	protected void executeStep(AnB_Session s, SessionState st,
	Carlsen_Steps step) {
		switch (step) {
			case STEP_1:
				st.VAR_INTR_R1 = (AnBx_Params)
				AnB_Session.readObject("Carlsen_1_STEP_1.ser");
				wffCheck("1.1 WFF", st.VAR_INTR_R1.getValue(1));
				eqCheck("1.2 EQ", aliases.get("ROLE_A"),
				st.VAR_INTR_R1.getValue(0));
				break;
			// ... remaining steps
		}
	}
}
	\end{lstlisting}
	\caption{Generated intruder code in reconstruction mode
	(\texttt{-objcheck}) for the Carlsen protocol. The step-file prefix
	(\texttt{Carlsen\_1\_}, \texttt{Carlsen\_2\_}, \dots, one per
	session) is baked into the generated source by the per-session
	compiler invocation with \texttt{-objchecksession <sid>}; this is
	why the pipeline regenerates the intruder role for every session
	inside Pass~B.}
	\label{fig:intr-code}
\end{figure}

\section{Experimental Results}
\label{sec:results}

To validate the practical viability of our passive intruder
reconstruction methodology, we conducted an evaluation on a test suite of
121 protocols. The evaluation covers Java-serialised traffic over TCP and
addresses five questions: whether the toolchain can extract serialised
objects from PCAPs and map them to protocol steps; whether the passive
intruder role validates the reconstructed traces against the protocol
specification; whether the defensive checks discriminate on semantic
content; whether the framework detects modifications of a valid
reconstruction; and whether the pipeline is efficient enough for
practical use. The first two questions are addressed in
Sections~\ref{subsubsec:single-session}
and~\ref{subsubsec:multi-session}; the last three, together with the
independence of the extraction stage from the compiler, are addressed in
the three additional experiments of Section~\ref{subsec:additional}. The
interpretation of these results, together with the limitations of the
approach, is presented in Section~\ref{sec:discussion}.

\subsection{Experimental Setup}
\label{subsec:experimental-setup}

All experiments were performed on a machine equipped with an Intel Core i7-4700HQ CPU at 2.40~GHz with 16~GB of RAM, running Windows~11. The toolchain comprises the \anbx{} compiler version 2026.09 (compiled with GHC 9.10.3), TShark (Wireshark) 4.4.8 for PCAP processing, Java OpenJDK 21.0.12 and Docker Desktop 4.89 for executing the generated protocol code, Python 3.14.3 for the extraction scripts, and Apache Ant 1.10.17 for build automation.

The reconstruction pipeline was executed using the \texttt{run\_test.py}
script, which orchestrates each stage and collects timing data
automatically. The single-session experiment
(Section~\ref{subsubsec:single-session}) runs each protocol with one
execution of each agent role. A multi-session experiment, reported in
Section~\ref{subsubsec:multi-session}, additionally runs the 112 protocols
whose generated implementations support parallel sessions at \(N=2\) and \(N=3\) parallel sessions,
where \(N\) is the number of parallel protocol sessions; the remaining
9 protocols are reconstructed as single-session captures.
PCAPs were generated by executing the protocols with the passive
intruder variant of the generated Java code
(\texttt{-out:JavaDocker -jdockerpcap}). For each protocol,
\texttt{run\_test.py} orchestrates the full pipeline (Stages~1--5 of
Figure~\ref{fig:pipeline-overview}), from generating the Dockerised Java
code and capturing the PCAP to building sessions, running \texttt{ObjCheck}
per candidate, and finally executing the passive intruder to score and
select the winning arrangement. Timing data is recorded for each stage and
appended to a single CSV file with one row per run.

Generative AI tools were used during development to debug the Python
scripts implementing the reconstruction pipeline and the analysis
scripts used to aggregate the reported results, and to make minor code
adaptations in those scripts. All such output was reviewed and tested
against the pipeline's recorded outputs; the author verified the final
versions of all scripts and takes full responsibility for the
correctness of the code and of the results reported in this paper.

\subsection{Test Suite}
\label{subsec:test-suite}

The test suite comprises 121 protocols drawn from the standard \anbx{}
distribution, including a range of industrial payment protocols. 
The protocols represent a broad spectrum of security goals, including
secrecy, strong authentication, and weak authentication. They also cover a
range of cryptographic primitives, such as asymmetric encryption, symmetric
encryption, digital signatures, hash functions and HMAC, and Diffie-Hellman
key agreement. The test suite includes several well-known protocols from the literature:
the Needham-Schroeder Public Key protocol~\cite{DBLP:journals/cacm/NeedhamS78},
Woo-Lam~\cite{DBLP:journals/sigops/WooL94}, Carlsen~\cite{carlsen}, Andrew
Secure RPC~\cite{DBLP:journals/tocs/Satyanarayanan89},
TMN~\cite{DBLP:conf/crypto/TatebayashiMN89}, ISO symmetric key
protocols~\cite{isosymkey}, Diffie-Hellman authenticated
variants~\cite{jisa2016}, and protocols from the
SET~\cite{DBLP:journals/jar/BellaMP06} and
iKP~\cite{DBLP:journals/jsac/BellareGHHKSTHW00} families.

For all experiments, the following cryptographic configuration was used:
asymmetric encryption and signatures employed RSA with 2048-bit keys, using
the transformation \texttt{RSA/ECB/OAEPWithSHA-256AndMGF1Padding} and
signature algorithm \texttt{SHA256withRSA}; symmetric encryption used AES
with 256-bit keys in GCM mode (\texttt{AES/GCM/NoPadding}); key agreement
used Diffie-Hellman with a 4096-bit modulus; hash functions used
\texttt{SHA-256}; and HMAC used \texttt{HmacSHA256} with 512-bit keys. The
protocol code was generated using this configuration for both the standard JDK 21 cryptographic provider and the Bouncy Castle provider
(version 1.85.2), allowing us to compare the impact of the security provider
on reconstruction success.

\subsection{Metrics}
\label{subsec:metrics}

We consider two types of metrics: reconstruction accuracy and runtime
efficiency. Reconstruction accuracy captures how faithfully the pipeline
recovers objects and passes defensive checks, while runtime efficiency
captures the cost of each stage of the pipeline.

\paragraph{Reconstruction accuracy}
For each protocol we record the number of expected serialised objects
(corresponding to the protocol steps) and the number successfully recovered
by the pipeline. We also record the number of defensive checks executed by
the generated intruder and the number that pass. Four check types are
reported, all derived from the protocol specification: well-formedness
(\texttt{wff}), equality (\texttt{eq}), inversion (\texttt{inv}) and seen
(\texttt{seen}) checks. These checks validate that the reconstructed objects
match the expected types, values and freshness requirements defined in the
specification. They do not imply that the protocol is secure or vulnerable;
they confirm that the captured trace is consistent with the protocol's
expectations.

From these per-protocol measurements we derive four aggregate indicators,
reported in the results tables: \textit{ObjCheck completion} is the
fraction of runs for which the object checker produced at least one step
file; \textit{complete reconstruction} is the fraction of protocols for
which every expected step file was produced
(Definition~\ref{def:complete}); \textit{DY-consistent reconstruction}
is the fraction of protocols that ran the intruder role without any
failed check (Definition~\ref{def:dy-consistent}); and
\textit{successful reconstruction} is the fraction of protocols
satisfying both conditions (Definition~\ref{def:successful}). Percentages
of recovered objects and of passed checks are reported alongside the
absolute counts.

The reconstruction accuracy metrics are extracted automatically from the
log files using the \texttt{parse\_logs.py} script.

\paragraph{Runtime efficiency}
We measure the wall-clock time of each stage of the reconstruction pipeline
using the \texttt{run\_test.py} script, which records timings with Python's
high-resolution \texttt{time.perf\_counter()}. The measured stages are:
JavaDocker generation (\texttt{anbxc -out:JavaDocker -jdockerpcap}), Docker
execution (protocol capture), PCAP to streams (TShark + \texttt{jpcap2ser.py}),
session build (\texttt{sessionbuild.py}), stream matching (\texttt{ObjCheck}),
Java code generation (\texttt{anbxc -out:Java -objcheck}), and Java execution
(reconstruction and validation). The script additionally measures the actual
protocol execution time by inspecting the first and last timestamps in the
PCAP, providing a runtime measurement of the protocol itself that is
independent of Docker orchestration. Timing data is collected in a single
CSV file with one row per protocol per run, enabling analysis across the
test suite.

\subsection{Single Session Reconstruction}
\label{subsubsec:single-session}

We evaluated the reconstruction pipeline using both the standard JDK
cryptographic provider and the Bouncy Castle provider. Table~\ref{tab:provider-comparison} summarises the
results for each provider. 

\begin{table}[ht]
	\centering
	\small
	\begin{tabular}{l|r|r}
		\toprule
		\textit{Metric} & \textit{Standard JDK} & \textit{Bouncy Castle} \\
		\midrule
		ObjCheck completion                       & 100.0\% (121/121) & 100.0\% (121/121) \\
		Complete reconstruction &  94.2\% (114/121) &  92.6\% (112/121) \\
		DY-consistent      &  98.3\% (119/121) &  98.3\% (119/121) \\
		Successful reconstruction & 94.2\% (114/121) & 92.6\% (112/121) \\
		\midrule
		Protocols with missing objects            & 7                 & 9 \\
		Protocols with failed checks              & 2                 & 2 \\
		\midrule
		Expected steps (objects)                  & 461               & 461 \\
		Object coverage      				      & 97.8\% (451/461)  & 97.0\% (447/461) \\
		Objects missing                           & 10 (2.2\%)        & 14 (3.0\%) \\
		\midrule
		Total consistency checks performed        & 948               & 922 \\
		DY check accuracy						  & 99.8\% (946/948)  & 99.8\% (920/922) \\
		Checks failed                             & 2 (0.2\%)         & 2 (0.2\%) \\
		Equality checks                           & 410               & 398 \\
		Well-formedness checks                    & 528               & 514 \\
		Inversion checks                          & 0                 & 0 \\
		Seen checks                               & 10                & 10 \\
		\bottomrule
	\end{tabular}
	\caption{Reconstruction outcomes with the Standard JDK and Bouncy Castle
		providers (121 protocols, \(N=1\)). Complete, DY-consistent, and
		successful reconstructions correspond to
		Definitions~\ref{def:complete}, \ref{def:dy-consistent}, and
		\ref{def:successful}. At \(N=1\) the complete and successful
		fractions coincide because no protocol in this suite recovers every object while
		failing a check; the two diverge at \(N \geq 2\)
		(Section~\ref{subsubsec:multi-session}).}
	\label{tab:provider-comparison}
\end{table}

The reconstruction pipeline achieved higher object recovery (97.8\% vs.\
97.0\%), higher successful reconstruction (94.2\% vs.\ 92.6\%), and equal
DY check accuracy (99.8\% vs.\ 99.8\%) when run with the standard JDK
provider compared to Bouncy Castle. Successful reconstruction is defined
as complete object recovery together with a clean passive-intruder
execution (Definition~\ref{def:successful}). ObjCheck completion and
DY-consistency were identical across both providers (100.0\% and 98.3\%,
respectively), so the difference in successful reconstruction is driven
entirely by the object-recovery stage.

The check-pass figures are conditional on the checks being executed:
because checks run only on recovered objects, missing objects reduce the
denominator of the pass rate as well as the numerator. The object
recovery figures are therefore the primary accuracy measure, and the
check-pass figures should be read as the conditional accuracy of the
checks given that they ran at all. This also explains why the total
number of checks differs between the two providers in
Table~\ref{tab:provider-comparison}: the provider that recovers fewer
objects also executes fewer checks.

One protocol, the Needham-Schroeder-Lowe (NSL) protocol, failed with the
standard JDK provider due to a serialisation artefact, whereas it succeeded
with Bouncy Castle. Specifically, with the standard JDK provider we
observed an additional byte \texttt{0x67} in the extracted stream; further
packet-level analysis is required to establish its precise origin. With
Bouncy Castle, the serialised object length differs, the additional byte
falls outside the extraction window, and the stream remains clean. The
\texttt{TLS} protocol failed under both providers, and the
\texttt{IKEv2DS} protocol failed only under Bouncy Castle; both are
discussed further below.

No inversion checks were required for this test suite because the protocol
compiler implements inversion as part of the well-formedness checks for
encrypted messages. The check types reported are those explicitly labelled
in the logs by the generated code.

The difference in missing objects between the two providers reveals a
consistent pattern. Six protocols failed to recover all objects under
\emph{both} providers: \texttt{DH\_auth} (missing 1 of 5 steps under both
providers), \texttt{Rev\_2KP} (missing 1 of 8), \texttt{SET\_Rev\_v2}
(missing 2 of 10), \texttt{ShareAgree} (missing 2 of 6),
\texttt{ShareAgree2} (missing 2 of 7), and \texttt{TLS} (missing 1 of 7).
One additional protocol failed only under the standard JDK provider:
\texttt{NSL} (missing 1 of 3). Three additional protocols failed only under
Bouncy Castle: \texttt{H530} (missing 1 of 5), \texttt{H530fix} (missing 2
of 5), and \texttt{IKEv2DS} (missing 2 of 5). Notably, the three BC-only
failures all involve Diffie-Hellman key exchange, suggesting a potential
serialisation issue with Bouncy Castle's handling of DH-related
cryptographic objects.

The failed checks followed a parallel pattern. The \texttt{TLS} protocol
had exactly one failed check under both providers (1 of 17 in each case), a
check that could not be satisfied because the
corresponding step was not recovered. The \texttt{NSL} protocol had one
failed check under the standard JDK provider (1 of 3), consistent with its
missing object. The \texttt{IKEv2DS} protocol had one failed check under
Bouncy Castle (1 of 8), again consistent with its missing object. No other
protocol had a failed check under either provider.

The cross-tabulation of object completeness and check success reveals a
consistent pattern across both providers. For the standard JDK provider,
114 protocols (94.2\%) had both all objects recovered and all checks
passing, 5 protocols (4.1\%) had missing objects but no failed checks, and
2 protocols (1.7\%) had both missing objects and failed checks. For the
Bouncy Castle provider, 112 protocols (92.6\%) had both all objects
recovered and all checks passing, 7 protocols (5.8\%) had missing objects
but no failed checks, and 2 protocols (1.7\%) had both missing objects and
failed checks. No protocol with a failed check recovered every object.

The missing objects are consistent with the block-data fragmentation described in Section~\ref{subsec:java-serialisation}. When multiple objects are written to the same stream, these markers can fragment the stream, preventing complete recovery of all objects.

The provider-specific differences observed in our experiments highlight an
important practical consideration for passive intruder reconstruction: the
cryptographic provider is determined by the protocol implementation being
captured, not by the analyst. In real-world scenarios, the analyst have no
control over which provider a target implementation uses; they must work
with whatever provider is present. 

\paragraph{Performance Results}

Table~\ref{tab:performance} presents the average execution times for each
stage of the pipeline, collected automatically by \texttt{run\_test.py}.
The timing data is derived from 121 protocol runs for each provider, both
of which achieved 100\% ObjCheck completion. Two features of the JavaDocker
generation row require comment. First, the stage is bimodal: 60 of the 121
protocols generate in under 310ms, while the remaining 61 take between
1,050ms and 2,400ms. For the standard JDK provider this means the mean
(726ms) lies below the median (1,050ms), unlike any other row in the
table; the Bouncy Castle run does not show this pattern. Second, JavaDocker
generation (Stage~1) is provider-independent: the compiler invocation
carries no \texttt{-securityProvider} flag, and the provider is selected
only at Java code generation (Stage~5). Since Stage~1 does not select a
provider, the two values in that row cannot reflect a provider effect; the
difference between them is run-to-run variability between the two capture
campaigns.

\begin{table}[ht]
	\centering
	\footnotesize
	\setlength{\tabcolsep}{3pt}
	\begin{tabular}{l|l|rrr|rrr}
		\toprule
		& & \multicolumn{3}{c|}{\textit{Standard JDK}} & \multicolumn{3}{c}{\textit{Bouncy Castle}} \\
		\cmidrule(lr){3-5} \cmidrule(lr){6-8}
		\textit{Stage} & \textit{Operation} & \textit{Mean} & \textit{Med.} & \textit{Std} & \textit{Mean} & \textit{Med.} & \textit{Std} \\
		\midrule
		1 & JavaDocker generation          &   726 & 1{,}050 &   489 & 1{,}193 & 1{,}133 &   194 \\
		2 & Docker execution (capture)     & 41{,}553 & 40{,}512 & 2{,}397 & 46{,}128 & 45{,}241 & 4{,}125 \\
		3 & PCAP to streams                &   754 &   719 &   132 &   780 &   728 &   157 \\
		-- & Protocol execution (from PCAP) &   733 &   477 &   710 &   881 &   661 &   843 \\
		4 & Session build                  &   262 &   250 &    43 &   270 &   253 &    47 \\
		5a & Stream matching (ObjCheck)      & 3{,}557 & 3{,}405 &   667 & 3{,}667 & 3{,}432 &   655 \\
		5b & Java code generation           & 1{,}693 & 1{,}012 & 4{,}867 & 1{,}717 & 1{,}007 & 4{,}873 \\
		5b & Java execution (validation)    & 4{,}952 & 4{,}704 &   866 & 6{,}840 & 6{,}508 & 1{,}057 \\
		\midrule
		& \textit{Total} & \textit{54{,}230} & \textit{52{,}030} & \textit{7{,}163} & \textit{61{,}476} & \textit{59{,}058} & \textit{8{,}359} \\
		\bottomrule
	\end{tabular}
	\caption{Average execution times by pipeline stage for both cryptographic
		providers (121 protocol runs each, single session). All times in
		milliseconds. The total row reports the sum of the visible stage means. It may differ 		by a few milliseconds from the aggregated total recorded by run\_test.py, because the displayed stage means are individually rounded.}
	\label{tab:performance}
\end{table}

The reconstruction pipeline completes in approximately 61.5 seconds on
average with Bouncy Castle and 54.2 seconds with the standard JDK provider,
both for a single-session run.  The actual protocol execution time, measured directly from
the communication time in the PCAP, is only 1.4\% of the total in both
cases (mean 881ms for BC, 733ms for JDK; medians 661ms and 477ms,
respectively). This confirms that the dominant cost is container
orchestration, capture setup, and the timeout used to ensure all containers
terminate gracefully, rather than the reconstruction itself.

Excluding the Docker stage, the reconstruction-specific cost (the sum of JavaDocker generation, PCAP extraction, ObjCheck, Java code generation, and Java execution) is approximately 14.2 seconds for BC and 11.7 seconds for JDK, concentrated in Java execution (6.8s BC, 5.0s JDK) followed by ObjCheck (3.7s BC, 3.6s JDK); PCAP extraction, JavaDocker generation, and Java code generation each contribute under 3\% of the single-session-equivalent total. The 7.2-second difference in total time between the two runs is dominated by the Docker stage (46.1s BC vs.\ 41.6s JDK), whose standard deviation across runs (4.1s BC, 2.4s JDK) is of the same order as the provider-to-provider difference. Since Docker orchestration is provider-independent, this difference reflects run-to-run variability in container startup and shutdown rather than a genuine performance advantage; the residual gap in the reconstruction stages is concentrated in Java
execution and reflects the higher cryptographic execution cost of the Bouncy Castle provider in this configuration. The experiment establishes the observed cost difference; it does not isolate its cause.

\subsection{Multi-Session Reconstruction}
\label{subsubsec:multi-session}

We evaluated the pipeline at N=1, N=2 and N=3 on the full 121-protocol suite
using the standard JDK provider. The session-completion markers emitted by the
generated implementations at \(N \geq 2\) classify 112 protocols as
\emph{parallel-capable} (every honest role reports
\texttt{Parallel}/\texttt{Multithreaded} and
\texttt{SessionsCompleted == SessionsDeclared}) and 9 as
\emph{sequential-only} (every honest role reports \texttt{Sequential}, so the
implementation cannot run parallel sessions even when
\texttt{-jparallelclients N} is passed). We report the two populations
separately, because only the parallel-capable subset exercises the candidate
enumeration and selection machinery in Section~\ref{subsec:candidate-scoring}.

Two representative excerpts from the runtime logs are shown below.

\begin{lstlisting}[basicstyle=\scriptsize\ttfamily,breaklines=true]
	# Parallel-capable (every honest role reports a parallel marker)
	[ROLE_B-ALWAYS] - All sessions completed: 3/3 - Multithreaded Server
	[ROLE_A-ALWAYS] - All sessions completed: 3/3 - Parallel clients: 3
	
	# Sequential-only (every honest role reports a sequential marker)
	[ROLE_B-ALWAYS] - All sessions completed: 3/3 - Sequential
	[ROLE_A-ALWAYS] - All sessions completed: 3/3 - Sequential
\end{lstlisting}

The classification is drawn from the honest roles only. During capture, the
passive intruder is realised by the \texttt{DY} container, which performs
packet capture but executes no protocol logic; the generated intruder role
is used only during reconstruction
(Section~\ref{subsec:intruder-generation}), where it reads the per-step
files sequentially and does not participate in the capture.

\paragraph{Reconstruction success}
Table~\ref{tab:multi-session-jdk} summarises the results. The per-object
recovery rate is essentially flat across N (97.8\% at N=1, 97.1--97.2\% at
N=2 and N=3), and the DY check accuracy is also flat (99.8\%, 99.4\%,
99.6\%). The per-protocol success rate drops from 94.2\% at N=1 to 89.3\% at both
N=2 and N=3. Two factors contribute. First, a protocol is counted as failed
if \emph{any} of its sessions misses an object, so a wider set of protocols
is exposed to the serialisation artefact as the number of sessions grows:
the per-object recovery rate is stable across N, but the protocol-level
rate is not. Second, the cross-session freshness failures discussed below
affect three additional protocols at N=2 (\texttt{Orig\_1KP},
\texttt{Orig\_1KP\_Fixed} and \texttt{Orig\_2KP}), all of which recover every
object, and one at N=3 (\texttt{Orig\_1KP\_Fixed}). The equality of the
N=2 and N=3 success rates is therefore coincidental: the two session counts
fail on partially different sets of protocols.

\begin{table}[ht]
	\centering
	\small
	\begin{tabular}{l|r|r|r}
		\toprule
		\textit{Metric} & \textit{N=1} & \textit{N=2} & \textit{N=3} \\
		\midrule
		Protocols in scope              & 121 & 121 & 121 \\
		Parallel-capable                & 112 & 112 & 112 \\
		Sequential-only                 &   9 &   9 &   9 \\
		\midrule
		ObjCheck completion                            & 100.0\% & 100.0\% & 100.0\% \\
		Complete (all)                                 &  94.2\% &  91.7\% &  90.1\% \\
		Complete (parallel)                            &    --   &  92.0\% &  90.2\% \\
		DY-consistent (all)                            &  98.3\% &  95.0\% &  96.7\% \\
		DY-consistent (parallel)                       &    --   &  95.5\% &  97.3\% \\
		\textit{Successful (all)}         & 94.2\% & 89.3\% & 89.3\% \\
		\textit{Successful (parallel)}    & --     & 89.3\% & 89.3\% \\
		\textit{Successful (sequential)}  & 88.9\% & 88.9\% & 88.9\% \\
		\midrule
		Expected steps (all)            &  461 &  868 & 1275 \\
		Objects recovered (all)         & 451 (97.8\%) & 844 (97.2\%) & 1239 (97.2\%) \\
		Objects recovered (parallel)    &    --   & 791 (97.2\%) & 1186 (97.1\%) \\
		\midrule
		Total consistency checks (all)  &  948 & 1745 & 2558 \\
		Checks passed (all)             & 946 (99.8\%) & 1734 (99.4\%) & 2548 (99.6\%) \\
		Checks passed (parallel)        &    --   & 1617 (99.4\%) & 2431 (99.6\%) \\
		\bottomrule
	\end{tabular}
	\caption{Multi-session reconstruction success with the standard JDK
		provider. Rows labelled \emph{all} cover the full 121-protocol suite,
		\emph{parallel} the 112 parallel-capable protocols, and \emph{sequential}
		the 9 sequential-only protocols. Complete, DY-consistent, and successful reconstructions correspond to
		Definitions~\ref{def:complete}, \ref{def:dy-consistent}, and
		\ref{def:successful}.}
	\label{tab:multi-session-jdk}
\end{table}

\paragraph{Candidate selection is necessary}
Table~\ref{tab:candidate-selection} quantifies how often the time-based
heuristic used during connection pairing would have produced a lower-scoring
arrangement than the winner. At N=2, 9 of the 51 K=2 protocols (17.6\%)
selected a candidate other than the identity arrangement; at N=3, 13 of the
51 K=2 protocols (25.5\%). In most of these runs the winner passed every check while the
identity arrangement did not; in a small number of iKP-family protocols the
winner had strictly fewer failed checks but did not eliminate them entirely.
For \texttt{Orig\_1KP}, \texttt{Orig\_1KP\_Fixed} and \texttt{Orig\_2KP} at
N=2, and for \texttt{Orig\_1KP\_Fixed} at N=3, the residual failures are
cross-session freshness checks and all objects are recovered. For
\texttt{Orig\_2KP\_Fixed} at both session counts, in contrast, the residual failures follow from the missing objects described under
\emph{Failure modes specific to multi-session} below. The score table in Section~\ref{subsec:candidate-scoring} is therefore
necessary: the time-based heuristic alone would have produced a
lower-scoring arrangement in a non-trivial fraction of the K=2 protocols
at both session counts.

\begin{table}[ht]
	\centering
	\small
	\begin{tabular}{l|r|r}
		\toprule
		\textit{Selection behaviour} & \textit{N=2} & \textit{N=3} \\
		\midrule
		Protocols with $\mathit{MaxCandidates} > 1$ & 53 & 53 \\
		\quad of which $K=2$ (2 and 6 candidates)   & 51 & 51 \\
		\quad of which $K=3$ (4 and 36 candidates)  &  2 &  2 \\
		Winner $\neq$ candidate~0                  &  9 & 13 \\
		\quad as \% of $K=2$ protocols             & 17.6\% & 25.5\% \\
		\quad as \% of all protocols               &  7.4\% & 10.7\% \\
		\bottomrule
	\end{tabular}
	\caption{Candidate selection behaviour. At N=3, roughly one in four
		K=2 protocols requires a non-identity arrangement. Candidate~0 is
		the time-based ordering; a non-zero winner means the time-based heuristic alone would have produced
		a reconstruction with a lower score than the winner; in most cases the
		winner passes every check, while in a small number of iKP-family protocols
		the winner has strictly fewer failed checks.}
	\label{tab:candidate-selection}
\end{table}

\paragraph{Failure modes specific to multi-session}
Two distinct failure patterns appear at N$\geq$2. The first is the extension
of the N=1 missing-object failures: every protocol that loses objects at N=1
continues to lose them at N=2 and N=3, and a small number of additional K=2
protocols join them (\texttt{Orig\_2KP\_Fixed}, \texttt{Orig\_3KP\_Fixed} and
\texttt{SET\_Orig\_v2} at both N=2 and N=3, and \texttt{Orig\_2KP} and
\texttt{Orig\_3KP} at N=3 only). The missing objects are of the same kind as the N=1 failures, consistent with the block-data fragmentation described in Section~\ref{subsec:java-serialisation}.
The number of missing objects per protocol varies across N and does not follow a
simple scaling law.

The second pattern is specific to N$\geq$2 and does not occur at N=1. At
N=2, three iKP-family protocols (\texttt{Orig\_1KP}, \texttt{Orig\_1KP\_Fixed}
and \texttt{Orig\_2KP}) recover every expected object yet still fail some
checks. The failures occur in checks that verify cross-session freshness
properties: when the candidate enumeration does not pair the matching
sessions across channels, the freshness check cannot be satisfied even though
all messages were received. For these three protocols at N=2, no candidate
arrangement passes every check, and the scoring loop selects the least-bad
one. At N=3, \texttt{Orig\_1KP\_Fixed} is the only protocol that continues to
show this pattern; \texttt{Orig\_1KP} recovers every object and has at
least one clean candidate, while \texttt{Orig\_2KP} moves to the
missing-object pattern described above and also has a clean candidate.
The failure is therefore not
attributable to concurrent execution per se, but to the inability of the
current enumeration to distinguish the matching sessions in these cases.

\paragraph{Performance scaling}
Table~\ref{tab:performance-scaling} reports the reconstruction cost as a
function of N, measured with the standard JDK provider on the full
121-protocol suite. The single-session-equivalent total grows from 54.2s
at N=1 to 60.9s at N=2 and 87.8s at N=3. Docker execution dominates at
every session count (41.6s, 43.1s and 44.0s respectively), and the
stages that do not scale with N (PCAP extraction, session build, and JavaDocker generation) are essentially flat across the three session counts. The
growth is concentrated in the three stages that scale with the candidate
space (ObjCheck, Java code generation and Java execution), because the
number of candidate arrangements grows with N: a K=2 protocol has a single
candidate at N=1, two at N=2 and six at N=3.

The means of these three stages are, however, skewed by a small number of
K$\geq$2 protocols; the medians tell a clearer story. The median ObjCheck
cost is 3.4s, 3.9s and 3.9s at N=1, N=2 and N=3 respectively; the
median Java code generation cost is 1.0s, 1.2s and 1.0s; the median
Java execution cost is 4.7s, 5.4s and 6.0s. For a typical protocol,
therefore, the per-session reconstruction cost is essentially flat across~N,
and the increase in the mean is driven by the tail (protocols with K=2 and
long message sequences, such as \texttt{Orig\_3KP\_Fixed}, which contribute a large fraction of the total at
N=3). The absolute wall-clock cost remains dominated by Docker orchestration
at every session count, as in the N=1 results of Section~\ref{subsubsec:single-session}.

\begin{table}[ht]
	\centering
	\small
	\setlength{\tabcolsep}{4pt}
	\begin{tabular}{l|l|r|r|r}
		\toprule
		\textit{Stage} & \textit{Operation} & \textit{N=1} & \textit{N=2} & \textit{N=3} \\
		\midrule
		1 & JavaDocker generation          &    726 &  1{,}090 &    781 \\
		2 & Docker execution (capture)     & 41{,}553 & 43{,}135 & 43{,}988 \\
		3 & PCAP to streams                &    754 &    798 &    773 \\
		-- & Protocol execution (from PCAP) &    733 &    637 &    854 \\
		4 & Session build                  &    262 &    259 &    261 \\
		5a & Stream matching (ObjCheck)    &  3{,}557 &  5{,}431 & 15{,}062 \\
		5b & Java code generation          &  1{,}693 &  2{,}537 &  7{,}363 \\
		5b & Java execution (validation)   &  4{,}952 &  7{,}009 & 18{,}673 \\
		\midrule
		& \textit{Total (single-session eq.)} & \textit{54{,}230} & \textit{60{,}896} & \textit{87{,}755} \\
		\bottomrule
	\end{tabular}
	\caption{Reconstruction cost as a function of N (standard JDK provider,
		121 protocol runs per session count, means in milliseconds). The
		JavaDocker, Docker, PCAP and session-build stages are independent of
		N and are included for completeness. ObjCheck, Java code generation
		and Java execution grow with the number of candidate arrangements:
		at N=3, a K=2 protocol has six candidates to reconstruct and score,
		against two at N=2 and one at N=1. The means of these three stages
		are therefore dominated by the K$\geq$2 protocols (see the medians
		quoted in the text), and should not be read as a per-session cost.}
	\label{tab:performance-scaling}
\end{table}

\subsection{Additional Validation}
\label{subsec:additional}

The main evaluation uses compiler-generated implementations and the \anbx{} specifications they were compiled from.
The three experiments below exercise aspects of the
pipeline that the main evaluation does not cover: extraction from
hand-written implementations (Section~\ref{subsubsec:handwritten}),
discrimination of the checks on semantic content
(Section~\ref{subsubsec:discrimination}), and detection of modifications to
a valid reconstruction (Section~\ref{subsubsec:mutations}).

\subsubsection{Application to Hand-Written Implementations}
\label{subsubsec:handwritten}

We conducted three additional experiments to determine whether the
reconstruction pipeline is specific to compiler-generated code, specific
to code that uses the \anbxj{} runtime library, or applicable to any Java
implementation that uses Java serialisation over TCP. In the three
experiments below, the object-to-step mapping that \texttt{ObjCheck}
performs in Stage~5, Pass~A (Section~\ref{subsec:objcheck-mapping}) was instead
performed by a hand-written extractor.

\paragraph{Experiment~1: Tutorial\_06}
A minimal two-agent program, using \anbxj{} exchanges five messages over a single TCP connection: three from server to client (\texttt{String}), one from client
to server (\texttt{String}), and one from client to server
(\texttt{Crypto\_SignedObject}). Both agents ran on localhost, so the two
directions share an IP address and the pairing stage distinguishes them
by port alone. Stage~3 recovered the two directional streams, the
hand-written extractor distributed the objects into five per-step files,
and the reported classes matched the sent classes on every step,
including \texttt{anbxj.Crypto\_SignedObject} on step~4.

\paragraph{Experiment~2: Tutorial\_07}
To isolate the dependence on the \anbxj{} library, we wrote a second
two-agent program using only the standard JDK, with no \anbxj{} jar on the
classpath. The exchange has the same five-step structure as Experiment~1,
with cryptographically wrapped objects on two steps: a
\texttt{javax.crypto.SealedObject} wrapping a symmetrically encrypted
\texttt{String} on step~3, and a \texttt{java.security.SignedObject}
wrapping a signed \texttt{String} on step~4. These are the underlying
types that \anbxj{} wraps in \texttt{Crypto\_SealedObject} and
\texttt{Crypto\_SignedObject}. Stage~3 and the extractor recovered all
five objects with the expected classes. The wrapped objects were read
back as opaque serialised values: the extractor requires no key material
and neither decrypts nor verifies them.

\paragraph{Experiment~3: Hybrid RSA+AES}
The third program, adapted from a public example of hybrid RSA+AES
encryption over TCP\footnote{\url{https://github.com/mark4865/JavaCryptoClientServer}},
probes two wire-format paths that neither of the first two experiments
exercises. It uses only the standard JDK. The exchange has four steps:
the server sends an RSA-2048 public key; the client generates an AES-256
session key, wraps it with the server's public key, and sends the wrapped
key; the server replies with the AES-encrypted message and a detached RSA
signature. The public-key step is interface-typed: the server writes a
\texttt{java.security.PublicKey}, which the JDK serialises as its internal
concrete class \texttt{sun.security.rsa.RSAPublicKeyImpl}. The remaining
three objects are raw \texttt{byte[]} arrays, serialised with the
\texttt{TC\_ARRAY} token. Stage~3 and the extractor recovered all four
objects with the expected classes.

\paragraph{Interpretation}
Together, the three programs show that the extraction stage depends only
on the wire format. \texttt{Tutorial\_06} was written by hand rather than
generated by the compiler and was recovered exactly, so object recovery
does not require the implementation to have been produced by the \anbx{}
compiler. \texttt{Tutorial\_07} uses only the standard JDK, with no
\anbxj{} jar on the classpath at any point, and was also recovered
exactly, so the recovery does not depend on the \anbxj{} runtime. The
hybrid program covers two further wire-format paths, an interface-typed
public key and raw \texttt{byte[]} arrays, and was recovered exactly as
well, so the recovery is not limited to objects whose static type is the
concrete class carried on the wire. Any Java program that puts
Java-serialised objects, including cryptographically wrapped ones, on a
TCP stream can therefore be reconstructed, given a channel-to-step map,
regardless of the library it uses, of whether it was compiler-generated,
and of the type of the objects on the wire. What the three experiments do
not do is exercise the compiler-generated checks or the
candidate-selection loop: the hand-written programs have no \anbx{}
specification from which those checks are derived, and the recovered
objects were therefore not validated against a type system, equality
constraints, or freshness requirements. A hand-written implementation can
be reconstructed without a formal model, given a channel-to-step map;
validating the reconstruction still requires one.

\subsubsection{Discrimination on a Role-Mapping Mismatch}
\label{subsubsec:discrimination}

The hand-written experiments do not exercise the defensive checks, so
they cannot show that the checks discriminate between role mappings.
We therefore constructed a pair of captures in which the wire structure
is held constant and only the mapping from roles to agent identities
varies.

The Carlsen specification was compiled twice with two different runtime
role mappings. In the first, the three roles are instantiated as
\texttt{alice}, \texttt{bob}, and \texttt{charlie}; in the second, as
\texttt{david}, \texttt{eve}, and \texttt{frank}. The two captures have
the same number of messages, the same types and arity at each step, and
the same cryptographic structure; they differ only in the identity
strings carried on the wire. The compiled intruder role is identical in
both cases, because the role-to-agent mapping is read from a properties
file at runtime rather than baked into the compiled code. We then ran
the intruder against each capture under each of the two check
instantiations, yielding a four-cell matrix.

Table~\ref{tab:discrimination} reports the outcome. The diagonal cells
pass every check. The off-diagonal cells fail at the intruder's equality checks on role
identities (steps 1.2, 2.2, 2.3). All other checks (including the
nonce equality check at step 2.4 and the well-formedness
checks) pass in every cell. No other types of checks are generated for
Carlsen. The failure is therefore specific to the identity comparison.

\begin{table}[ht]
	\centering
	\small
	\begin{tabular}{l|c|c}
		\toprule
		& \textit{checks from \textsf{ref}} & \textit{checks from \textsf{swap}} \\
		\midrule
		\textit{capture \textsf{ref}}  & PASS & FAIL (1.2, 2.2, 2.3) \\
		\textit{capture \textsf{swap}} & FAIL (1.2, 2.2, 2.3) & PASS \\
		\bottomrule
	\end{tabular}
	\caption{Role-mapping discrimination. The same compiled intruder is
		run against each capture under each check instantiation. Only
		the identity equality checks fire, and only in the off-diagonal
		cells.}
	\label{tab:discrimination}
\end{table}

The result is clear: the same capture that fails under one check
instantiation passes under the other. The failure is therefore
attributable to the mismatch between the check instantiation and the
role mapping that produced the capture, not to any incidental property
of either capture or of the compiled intruder. Together with the
hand-written experiments of Section~\ref{subsubsec:handwritten}, this
shows that the checks are not merely satisfied by well-formed messages;
they respond to a semantic property of the trace relative to the
specification instance against which the trace is validated.

\subsubsection{Detection of Mutations in Reconstructed Traces}
\label{subsubsec:mutations}

The role-mapping experiment of Section~\ref{subsubsec:discrimination}
shows that the checks reject a trace whose role mapping disagrees with
the check instantiation, but it does not show that the framework detects
arbitrary modifications of a valid reconstruction. Since the solution is
presented as a validator rather than a deserialiser, it must be shown
that modifications can be caught. We therefore applied three file-level
mutations to the winning reconstruction of each protocol and measured the
fraction detected by each of two runtime signals. The mutations act on
the per-step \texttt{.ser} files, not on the PCAP: \emph{deletion}
removes the step at index $\lfloor n/2 \rfloor$ of the first session
($n \geq 2$); \emph{duplication} copies the first step file of the first
session over the second ($n \geq 2$); and \emph{cross-session swap}
copies the step at index $\lfloor n/2 \rfloor$ of the second session
over the same position in the first ($N \geq 2$). A mutation is
\emph{flagged by check failure} if the passive intruder reports a failed
check or exits non-zero, and \emph{flagged by check-count deficit} if it
executes fewer checks than the unmutated baseline. The two categories
are mutually exclusive by attribution: a mutation that both fails a check
and suppresses a later check is attributed to the check failure.
Protocols whose unmutated baseline fails a check are excluded, because a
mutation on an already-flagged trace cannot be attributed to the
mutation. The check-count deficit is a runtime diagnostic and does not
evaluate the completeness predicate of Definition~\ref{def:complete}
directly; the two are related but not equivalent, as discussed below.

\paragraph{Results}
Table~\ref{tab:mutations} reports the outcome over twenty protocols at
N=1 and N=2. The twenty span the structural features the experiment
exercises ($K \in \{1,2,3\}$, 2 to 10 steps per session, 6 to 52 checks
per session), with one representative per protocol family to avoid
near-identical check structures; all twenty are parallel-capable at
N=2 and have a clean baseline at both session counts. Deletion is
flagged in 34 of 40 traces, duplication in 28 of 40, and cross-session
swap in 3 of 20, for a total of 65 of 100. Of those, 27 produce a check
failure or non-zero termination and 38 produce only a check-count
deficit; the remaining 35 mutations trigger neither signal. 
The attribution between the two signals is stable across N for every protocol.

\begin{table}[ht]
	\centering
	\small
	\setlength{\tabcolsep}{4pt}
	\begin{tabular}{l|c|r|>{\raggedleft\arraybackslash}p{2.2cm}|>{\raggedleft\arraybackslash}p{2.2cm}|r}
		\toprule
		\textit{Mutation} & \textit{$N$} & \textit{Traces} &
		\textit{By check failure} & \textit{By check-count deficit} &
		\textit{Flagged} \\
		\midrule
		Deletion            & 1, 2 & 40 &  8 & 26 & 34 \\
		Duplication         & 1, 2 & 40 & 16 & 12 & 28 \\
		Cross-session swap  & 2    & 20 &  3 &  0 &  3 \\
		\midrule
		Total               &      & 100 & 27 & 38 & 65 \\
		\bottomrule
	\end{tabular}
	\caption{Operational detection of mutations in reconstructed traces
		over twenty protocols. A trace is flagged \emph{by check failure}
		if the passive intruder reports a failed check or exits non-zero,
		and \emph{by check-count deficit} if it executes fewer checks than
		the unmutated baseline. The two columns are mutually exclusive by
		attribution: a mutation that both fails a check and suppresses a
		later check is attributed to the check failure. The \emph{Flagged}
		column is their union. A check-count deficit is a runtime diagnostic
		and does not evaluate the completeness predicate of
		Definition~\ref{def:complete} directly.}
	\label{tab:mutations}
\end{table}

\paragraph{Interpretation}
Two observations bound what the mutation experiment shows. First, the
two runtime signals are independent of the formal properties they most
resemble. Three protocols (SET\_Rev, Rev\_1KP, Rev\_3KP) accept deletion
without any runtime signal: the deleted step file is absent so the trace
is incomplete by Definition~\ref{def:complete}, yet the middle step
carries no checks that depend on its content, so the executed validation
activity is unchanged. The divergence also runs the other way: a check
failure aborts the current step and skips any later checks within it, so
a reconstruction with all step files present can produce a check-count
deficit, just as one with a missing step file can produce none. Second,
detection depends on whether the specification places a check whose
outcome depends on the mutated content, not on the protocol's step count,
its check count, or the number of parallel sessions. This explains why
the three longest protocols (SET\_Rev, Rev\_1KP, Rev\_3KP) are not
flagged by deletion, and why some protocols accept duplication: the
substituted content either satisfies the checks generated for the
affected step, or the step carries no content-dependent checks.
Cross-session swap is detected in three of twenty protocols
(Otway--Rees, SET\_Orig, H530), the three that expose values on the
affected step that the intruder can compare against session-specific
state; in the remaining seventeen the affected step carries symmetric
ciphertext or content whose value does not differ between the two
sessions. This is what the Dolev--Yao model predicts: a passive adversary
without the session key cannot distinguish two ciphertexts of the same
shape. The limit is on the intruder's information, not on the framework's
design.

\section{Discussion and Limitations}
\label{sec:discussion}

In this section we discuss the results presented in Section~\ref{sec:results} and analyse their limitations.

\subsection{Discussion}
\label{subsec:discussion}

\paragraph{What the toolchain demonstrates}
The passive Dolev--Yao intruder model can be instantiated in practice. A
passive adversary with knowledge of the protocol specification can extract
serialised Java objects from captured network traffic, reconstruct the
protocol state, and validate that the observed messages are well-formed
according to the protocol's type system and knowledge sets. This is the
threat model underlying bulk surveillance operations that record network
traffic for later analysis, and the toolchain provides a concrete mechanism
for understanding what such an adversary can derive from captured traffic.
It also forms a practical bridge between formal specifications and concrete
traffic: the defensive checks enable automated consistency checking of
captured traces against specifications, providing diagnostic information
that can guide further investigation. The mutation experiment of
Section~\ref{subsubsec:mutations} characterises the limits of this diagnostic
capability; the evidence is presented there and not repeated here.

\paragraph{What the checks can and cannot detect}
The accuracy figures of Section~\ref{subsubsec:single-session} conceal a
distinction that matters for interpreting the checks. At \(N=1\), every
check failure is a downstream consequence of a missing object: because
checks run only on recovered objects, a missing value either skips the
corresponding check or breaks a later equality that depends on it, and no
protocol with a failed check recovers every object. At \(N \geq 2\) the
iKP-family protocols break this pattern: they recover every expected object
and yet fail cross-session freshness checks, because no candidate
arrangement pairs the matching sessions across channels. The distinction
separates what the checks can detect from what they cannot: they detect
structural inconsistencies in recovered messages, but they do not resolve
an ambiguity in the session grouping. The latter is a limit of the
enumeration strategy, not of the checks themselves.

\paragraph{Cost and provider dependence}
The reconstruction-specific stages are inexpensive relative to capture and
storage (Section~\ref{subsubsec:single-session}), which makes the pipeline
suitable for interactive analysis and forensic investigation, though not
for real-time traffic analysis; real-time analysis is in any case not the
primary concern of a harvest now, decrypt later strategy, which focuses on
retrospective analysis of stored traffic. A related observation is that the
cryptographic provider is a property of the target implementation, not of
the analyst. Provider-specific failures should therefore be treated as artefacts of the implementation, not as evidence about the protocol.

\paragraph{Candidate selection is not a fixed property}
The score-based selection of Section~\ref{subsec:candidate-scoring} is
necessary, not merely an optimisation on the time-based heuristic
(Table~\ref{tab:candidate-selection}). More importantly, the winning
candidate is not fixed across session counts of the same protocol:
Carlsen selects candidate~1 at \(N=2\) and candidate~0 at
\(N=3\), and the arrangement that the heuristic produces at \(N=2\)
fails two checks while the same arrangement at \(N=3\) passes all of
them. We cannot attribute this to \(N\) itself, since the runs differ in
timing, in the number of connections, and in the candidate space. What
the observation establishes is that the winning candidate is not a static
property of the protocol alone, and that score-based selection is required
to identify the best-scoring arrangement for each individual
reconstruction.

\subsection{Limitations}
\label{subsec:limitations}

\paragraph{No decryption without keys}
Encrypted traffic cannot be decrypted without the appropriate
cryptographic keys. For protocols that provide forward secrecy, such as
TLS 1.3, a passive adversary cannot recover the plaintext even if
long-term keys are later compromised. Our toolchain is therefore best
suited for evaluating protocols that lack forward secrecy, or for
analysing traffic that has been decrypted through other means.

\paragraph{Formal-model limitations}
Our approach requires a formal specification of the protocol in the
\anbx{} language, from which the compiler derives both the defensive
checks and the channel-to-step map that Stage~5 uses. The toolchain can
therefore only be used for protocols that have been modelled and verified;
undocumented protocols would require manual reverse engineering. The
hand-written experiments of Section~\ref{subsubsec:handwritten} clarify
the boundary: the extraction stage can run on implementations the analyst
did not generate, but a channel-to-step map must still come from
somewhere, from the specification, as in this work, or from an analysis
of the implementation or of the captured trace. Even with a specification
in hand, the language supports stateless protocols and those with sequence
numbers, but not complex stateful behaviour such as session resumption or
key rotation; this would require extensions to both the language and the
pipeline. Finally, the analysis is symbolic, and does not account for
probabilistic or side-channel attacks.

\paragraph{Implementation-level limitations}
Three engineering limitations are worth noting. First, the toolchain
relies on deployment metadata, expressed in our implementation as the
\texttt{.properties} file. \texttt{sessionbuild.py} derives the number of
connections per session \(K\) from the channel structure, and the expected
channel order guides object-to-step assignment; a PCAP whose network
configuration does not match the file will fail during session build or
produce an incorrect channel assignment. Second, the candidate space for
multi-session reconstruction grows as \((N!)^{K-1}\), and
\texttt{sessionbuild.py} caps enumeration at \(120\) arrangements,
falling back to a heuristic set above the cap; this set is sufficient only
if the correct arrangement is among the enumerated candidates, which is
not guaranteed. Third, the scoring function is a heuristic lexicographic
tuple, and the pipeline reads its inputs from textual markers
(\texttt{CHECK \# <id>} and \texttt{check FAILED!}) emitted by the
\anbxj{} library rather than from a structured report interface. An
incorrect arrangement could pass every defensive check by coincidence and
tie with the correct one, and a change to the library's output format
would silently stop the parser from matching.

\paragraph{Java-specific implementation}
The current implementation is tied to Java: the extraction stage
(\texttt{outparse.py} and \texttt{ObjCheck}) parses Java's wire format,
and the generated code deserialises Java objects via
\texttt{ObjectInputStream}. The pipeline should therefore only be run on
PCAP files from trusted sources or within isolated environments, since
Java deserialisation vulnerabilities are well-documented~\cite{Sayar2023}.
The format dependency is confined to the extraction stage; the rest of
the pipeline operates on the extracted objects rather than on bytes.

\paragraph{Scope of evaluation}
The benchmark evaluates reconstruction under controlled captures generated
from known implementations and specifications. All captures are
self-generated: in all three experiments we ran the
implementations and captured the traffic, though in the third the code was
written independently. The evaluation therefore establishes technical
feasibility and pipeline robustness, but does not establish performance on
independently captured operational traffic. We are not aware of any public
corpus of captures of Java-serialised protocol executions against which
the pipeline could be validated; producing one would be a useful
contribution in its own right.

\section{Related Work}
\label{sec:related-work}

Our work is positioned at the intersection of several research areas:
global surveillance, attack trace reconstruction, model-based security
testing, protocol reverse engineering, passive adversary models, network
forensics, and code generation from formal models.

\subsection{Global Surveillance}

The surveillance infrastructures that instantiate the passive adversary
model have been analysed extensively in the geopolitical, economic, and
sociological literature. Zuboff~\cite{zuboff2019surveillance,Zuboff2022}
frames the corporate form as ``surveillance capitalism'', an economic
logic that commodifies human experience; Penney's survey of over 1{,}200
internet users~\cite{Penney2017} provides evidence of the resulting
``chilling effect'', and Sklansky~\cite{sklansky2014} terms the
propagation of this effect beyond direct targets the ``stultification
thesis'' of surveillance.

The operational feasibility of such passive surveillance has been
quantified by Blanco-Romero et al.~\cite{blancoromero2026}, who model the
economics of harvest now, decrypt later attacks against TLS 1.2, TLS 1.3,
QUIC, and SSH. Their analysis demonstrates that the storage costs of
retaining intercepted ciphertext are economically trivial for state-level
actors (a 1\% selective harvest of global traffic is estimated at
approximately \$1.1 billion annually, well within national intelligence
budgets) and that long-term retention on magnetic tape media incurs
near-zero operational cost over multi-decade timeframes. 
Where Blanco-Romero et al.\ quantify the economics of storing and
decrypting captured traffic, our reconstruction pipeline provides the
forensic mechanism to validate concrete protocol traces against formal
specifications.

\subsection{Attack Trace Reconstruction}

A direct precursor to our work is the automatic generation of attack
specifications and implementations from OFMC traces by Garcia and
Modesti~\cite{GARCIA2024100038}. That work introduced a compiler extension
that takes an OFMC attack trace and produces both an \anbx{} specification
that includes the intruder's actions and executable Java code for an
\emph{active} intruder that realises the attack. The extension is limited
to a single protocol session. Our present work extends this in three
significant directions. First, we operate on concrete network traffic
(PCAP files) rather than abstract attack traces, extracting the actual
serialised objects that correspond to the abstract messages. Second, we
model a \emph{passive} intruder that observes and validates captured
traffic, rather than an active one that manipulates it; the generated code
therefore reads from files rather than the network and performs defensive
checks on the reconstructed messages instead of executing an attack
strategy. Third, our pipeline handles multiple parallel sessions, whereas
attack trace reconstruction is limited to a single session. The two
approaches are complementary: attack trace reconstruction produces a
concrete attack implementation from a symbolic counterexample, whereas our
pipeline validates captured traffic against the specification.

\subsection{Model-Based Security Testing}

Several approaches have explored the bridge between formal models and
implementation testing. Sivelle et al.~\cite{sivelle2022} synthesise
implementations of secure protocols and attacks from abstract models,
compiling symbolic attacks to F* code; Ghabri et
al.~\cite{DBLP:journals/corr/GhabriMR13} compile symbolic attacks to
protocol implementation tests; and the model-based testing approaches of
Dadeau et al.~\cite{DBLP:journals/stvr/DadeauHKMR15} and Maatoug et
al.~\cite{maatoug:hal-01089682} generate tests from models but do not
operate on captured network traffic. Armando et
al.~\cite{DBLP:conf/tap/ArmandoPCMB12} and the SPaCIoS
Tool~\cite{DBLP:conf/icst/Vigano13} generate and execute security tests
from a formal description and a model of the attacker, and
SecureMDD~\cite{katkalov2012model} and
MobSTer~\cite{DBLP:journals/stvr/PeroliMVG18} derive test cases from
higher-level models. A more recent direction combines model learning with
formal analysis, extracting an implementation model and verifying it
against an Applied Pi calculus specification~\cite{Tian2025}; our
approach assumes the specification is known and focuses on the
reconstruction of concrete traffic rather than on model extraction.
Mutation-based testing, as in Dadeau et
al.~\cite{DBLP:journals/stvr/DadeauHKMR15}, generates negative test cases
by mutating the formal model; our mutation experiment of
Section~\ref{subsubsec:mutations} plays an analogous role at the trace
level. The work by Ammann et al.~\cite{Ammann2024} on DY Fuzzing combines
formal Dolev--Yao models with cryptographic protocol fuzz testing, but
focuses on active fuzzing rather than passive trace reconstruction.

\subsection{Protocol Reverse Engineering}

Protocol reverse engineering (PRE) infers specifications from traffic
or from implementations, whereas we validate traffic against a known
specification. The two are complementary: PRE tools could generate
candidate specifications for protocols that have not been formally
modelled, which our pipeline could then validate.
Most PRE tools operate on traffic. \textit{Netzob}~\cite{Netzob}
infers both message formats and state machines through passive and
active processes, and is representative of a family that also includes
\textit{MARISSA}~\cite{MARISSA2026} and
\textit{Discoverer}~\cite{Cui2007}, which recover message formats by
similarity-based clustering and type-based sequence alignment. A second
group focuses on state machines rather than formats, including
\textit{IPSMInfer}~\cite{IPSMInfer2025} for industrial protocols and
\textit{StruFSM}~\cite{StruFSM2026} at the byte level. A third family
analyses the implementation rather than the traffic alone:
\textit{Polyglot}~\cite{Caballero2007} and
\textit{Prospex}~\cite{Comparetti2009} track data flows through memory
to recover field boundaries, and AUTHSCAN~\cite{DBLP:conf/ndss/BaiLMVS00D13}
recovers web authentication protocols from implementations. Our approach
provides a different and stronger form of guarantee, because it works
from a verified formal model rather than inferring one, but it requires
the specification to be available.

\subsection{Passive Adversary Models}

The Dolev--Yao model~\cite{dolev83ieee} provides the standard symbolic
formalism for analysing cryptographic protocols. While the active intruder
has received extensive attention in formal verification, the passive
adversary has been studied in fewer contexts.
Delaune and Jacquemard~\cite{DBLP:journals/jar/DelauneJ06} extend the
Dolev--Yao model to capture offline dictionary attacks, augmenting the
deduction system with probabilistic encryption and guessing, and establish
that the intruder deduction problem in this extended model is decidable in
PTIME. Baudet et al.~\cite{DBLP:journals/iandc/BaudetCK09} develop a general
framework for comparing formal and computational models in the presence of
passive adversaries. These works formalise what a passive adversary can
deduce from observed messages and when it can distinguish between two sets
of observations, the core capabilities of a passive intruder. Our work
addresses a different question: what a passive adversary can reconstruct
at the application protocol level, given knowledge of the specification.
Rather than solving symbolic constraints over observed terms, we extract
and validate the actual protocol messages from captured PCAP files,
performing the defensive checks derived by the compiler from the
specification.

\subsection{Network Forensics and Traffic Analysis}

Network forensics tools such as Wireshark, its command-line counterpart
TShark~\cite{tshark}, Tcpdump~\cite{tcpdump}, and Zeek~\cite{zeek}
capture and analyse network traffic, reconstruct flows, extract
application-layer protocols, and perform deep packet inspection.
Wireshark's ``Export Objects'' feature supports HTTP and SMB, and
NetworkMiner~\cite{networkminer} extracts files and images while
performing passive OS fingerprinting.

TShark is particularly relevant to our work, as it enables scriptable
extraction of packet data. We use its \texttt{tcp.stream},
\texttt{ip.addr}, and \texttt{data} fields to identify TCP streams and
extract application-layer payloads. Zeek is representative of the more
capable end of this spectrum: it passively analyses traffic and generates
structured logs for over 40 application-layer protocols, but its analysis
is limited to known protocols with established dissectors, and it cannot
inspect encrypted payloads or validate protocol semantics against a formal
specification. Despite these capabilities, network forensics tools operate
without semantic understanding of security protocols: they can reconstruct
TCP streams and display message bytes, but they cannot determine whether a
captured sequence of messages satisfies a security goal such as
authentication or secrecy. Our toolchain extends these tools by providing
protocol-level reconstruction and security validation. Where Wireshark can
show the raw bytes of a message, our tool can determine whether those
bytes, when deserialised, form a well-formed protocol message consistent
with the specification. The \texttt{jpcap2ser.py} and \texttt{outparse.py}
scripts provide a generic mechanism for extracting serialised Java objects
from TCP streams, and the generated intruder code uses the protocol
specification to validate the reconstructed trace against the expected
sequence of steps.

\subsection{Code Generation from Formal Models}

Several tools generate implementations from formal protocol models.
\textit{JavaSPI}~\cite{Sisto2018} generates Java code from Spi Calculus
specifications with a formal soundness proof;
\textit{Expi2Java}~\cite{backes2012development_expi2java} generates Java
code from extended Spi Calculus models; Meng et
al.~\cite{meng2017automatic} generate Java code from CryptoVerif models;
and \textit{Modular Choreographies}~\cite{CruzFilipe2023} compiles
Alice \& Bob communication syntax to Java via the Choral language. Recent
work on verified compilers for security protocols, such as
\textit{Noise*}~\cite{Ho2022} and \textit{Molly}~\cite{Dougherty2023},
provides machine-checked correctness proofs for specific protocol
frameworks, and Arquint et al.~\cite{Arquint2023, Arquint2023a} present a
generic methodology for modular verification of security protocol
implementations.

The \anbx{} compiler~\cite{anbx2015, jisa2016} generates Java code from
\anb{}/\anbx{} specifications, deriving defensive checks from the
protocol's knowledge sets; these checks operationalise the analysis and
synthesis operations of the symbolic Dolev--Yao model on concrete
messages. We extend this compiler with a reconstruction pipeline that
operates on captured traffic rather than on a specification alone, and
the resulting work differs from other code generation approaches in two
ways. First, our primary contribution is reconstruction from captured
traffic rather than code generation from a specification; the code
generation is a means to replay and validate the captured trace. Second,
our generated code implements a passive intruder that reads from files
rather than the network, together with the \texttt{ObjCheck} class that
extracts individual objects from the captured streams.
Table~\ref{tab:related-comparison} summarises the resulting differences.

\subsection{Comparison of Approaches}

\begin{table}[ht]
	\centering
	\footnotesize
	\begin{tabular}{p{3.6cm}|p{2.5cm}|p{2.5cm}|>{\centering\arraybackslash}p{0.7cm}|>{\centering\arraybackslash}p{0.7cm}}
		\toprule
		\textit{Approach} & \textit{Input} & \textit{Output} & \textit{Exec} & \textit{PCAP} \\
		\midrule
		OFMC~\cite{basin2005ofmc}, ProVerif~\cite{blanchet2016modeling}, Tamarin~\cite{tamarin2012automated} & Symbolic model & Abstract trace & -- & -- \\
		Attack trace reconstruction~\cite{GARCIA2024100038} & OFMC trace & \anbx{} specification and Java code (active) & \checkmark & -- \\
		Protocol reverse engineering (Netzob, IPSMInfer, etc.) & PCAP & Inferred protocol model & -- & \checkmark \\
		Network forensics (Wireshark, Zeek, etc.) & PCAP & Packets/flows & -- & \checkmark \\
		JavaSPI~\cite{Sisto2018}, Expi2Java~\cite{backes2012development_expi2java} & Spi Calculus & Java code & \checkmark & -- \\
		Modular Choreographies~\cite{CruzFilipe2023} & Alice \& Bob & Java code & \checkmark & -- \\
		Model-based security testing (SPaCIoS~\cite{DBLP:conf/icst/Vigano13}, SecureMDD~\cite{katkalov2012model}, MobSTer~\cite{DBLP:journals/stvr/PeroliMVG18}) & Model + goals & Tests / attacks & \checkmark & -- \\
		Model learning and fuzzing (Tian~\cite{Tian2025}, Ammann~\cite{Ammann2024}) & Model / implementation & Verified model / test inputs & \checkmark & -- \\
		\textit{Our work} & \textit{PCAP + \anbx{}} & \textit{Java intruder (passive) and ObjCheck} & \checkmark & \checkmark \\
		\bottomrule
	\end{tabular}
	\caption{Comparison of related work on formal analysis, protocol reconstruction, and validation.}
	\label{tab:related-comparison}
\end{table}

Table~\ref{tab:related-comparison} summarises the key differences between
our approach and existing tools. We are not aware of an existing tool that
combines all three: PCAP input, executable output, and validation against
a formal specification. The closest prior work is the attack trace
reconstruction of Garcia and Modesti~\cite{GARCIA2024100038}, which we
extend by taking captured traffic as input, extracting the serialised
objects from the PCAP, validating the reconstruction against the protocol
specification through the generated \texttt{ObjCheck} class and passive
intruder role, and handling multiple parallel sessions rather than a
single one. Our approach fills a gap by providing a link between formal
verification and operational security analysis.

\section{Conclusion and Future Work}
\label{sec:conclusion}

We presented an automated, specification-guided framework for
reconstructing passive Dolev--Yao intruder executions from captured
network traffic. The framework takes a PCAP capture produced under a
passive man-in-the-middle configuration and a protocol specification in
\anbx{}, and generates executable Java code that reconstructs the captured
trace and evaluates the defensive checks derived by the compiler from the
specification. The generated intruder does not modify traffic; it
observes, reconstructs, and validates. The framework's contributions are set out in Section~\ref{sec:introduction} and substantiated in
Sections~\ref{sec:methodology}--\ref{sec:results}. In brief, it provides
an operational instantiation of a passive Dolev--Yao intruder, a
PCAP-to-protocol reconstruction pipeline, specification-guided candidate
selection using Dolev--Yao-derived consistency checks, and an evaluation
over 121 protocols under multiple sessions, alternative cryptographic
providers, third-party Java traffic, role-mapping mismatches, and
controlled mutations. The toolchain is a proof-of-concept laboratory
model of one stage of passive surveillance analysis.

The evaluation shows that the approach is feasible in practice. With the
standard JDK provider, the pipeline achieves 94.2\% successful
reconstruction at \(N=1\) and 89.3\% on the 112 parallel-capable protocols
at \(N=2\) and \(N=3\), while recovering approximately 97\% of expected
objects. The extraction stage is independent of the \anbx{} compiler and
runtime library, and the defensive checks respond to semantic mismatches
and to several classes of mutation. These results establish a concrete link
between formal protocol specifications and concrete network traffic: the
checks are derived from the specification, not the implementation, and can
reject traces that are structurally well formed but inconsistent with the
specification instance.

The significance of the work lies in bridging formal verification and
practical network analysis. Verification tools operate on abstract models
and produce symbolic traces; network forensics tools display packets but
lack protocol semantics. The toolchain uses a formal specification to
interpret captured traffic, while remaining independent of whether
encrypted payloads can be decrypted. In a harvest now, decrypt later
setting, this means the protocol-level structure of a captured execution
can be recovered and validated before keys become available. 

Several directions remain for future work. Extending the \anbx{} compiler
with post-quantum primitives would allow the pipeline to be applied to
traffic whose long-term confidentiality is threatened by retrospective
quantum decryption. Implementing the extraction stage for Protocol Buffers
would broaden the toolchain beyond Java serialisation. A further possibility
is to reduce the pipeline's dependence on externally supplied deployment
metadata: both the \texttt{.properties} file and the channel-to-step map
used in the hand-written experiments could be inferred from the
implementation or from the observed message sequence, with candidate
mappings validated against the specification. This would make the
toolchain self-contained given only the specification and the PCAP.
Finally, content-based pruning guided by reconstructed nonces would help
scale multi-session reconstruction beyond the current enumeration cap and
address the cross-session freshness failures noted in
Section~\ref{subsubsec:multi-session}.

\paragraph*{Data Availability}
The source code and documentation for the \anbx{} compiler is available at  \url{https://paolo.science/anbx/} and the
reconstruction pipeline and the experiments are available at \url{https://github.com/paolo-mo/passive-dolev-yao-intruder}.

\paragraph*{Acknowledgement} The author would like to thank Rémi Garcia and Louis Holmes for insightful discussions. Furthermore, Louis ran preliminary experiments during his BSc dissertation, which helped to better understand some aspects of Java serialisation and stream extraction from PCAP files.

\paragraph*{Declaration of generative AI and AI-assisted technologies in the writing process}

During the preparation of this work the author used DeepSeek and ChatGPT
to check grammar and improve the readability of the manuscript. After
using these tool, the author reviewed and edited the content as needed
and takes full responsibility for the content of the published article.

\bibliographystyle{elsarticle-num}
\bibliography{literature}

\end{document}